\documentclass[10pt]{book}
\usepackage{sussp_TnF} 
\usepackage{makeidx}
\usepackage{epsfig,graphicx}
\usepackage{amssymb}
\usepackage{cite}
\makeindex

\newcommand{\be}{\begin{equation}}
\newcommand{\ee}{\end{equation}}
\newcommand{\ba}{\begin{array}}
\newcommand{\ea}{\end{array}}
\newcommand{\bea}{\begin{eqnarray}}
\newcommand{\eea}{\end{eqnarray}}

\newcommand{\cA}{{\cal A}}
\newcommand{\cO}{{\cal O}}
\newcommand{\cB}{{\cal B}}

\newcommand{\cL}{{\cal L}}

\newcommand{\cT}{{\cal T}}
\newcommand{\cM}{{\cal M}}

\newcommand{\no}{\nonumber}

\newcommand{\sla}{\! \! \! \!  /~}
\newcommand{\rhob}{\bar\varrho}
\newcommand{\etab}{\bar\eta}
\renewcommand{\Im}{{\rm Im}\,}
\renewcommand{\Re}{{\rm Re}\,}

\begin{document}

\headings{$B$ physics in the LHC era} 
{$B$ physics}
{Gino Isidori}
{INFN, Laboratori Nazionali di Frascati, I-00044 Frascati, Italy\\
 Institute for Advanced Study, 
 T.~U.~M\"unchen, D-80333 M\"unchen, Germany}

\section{Introduction}

In the last few years there has been a great experimental progress in quark 
flavour physics. The validity of the Standard Model (SM) has been strongly 
reinforced by a series of challenging experimental 
tests in $B$, $D$, and $K$ decays.  All the relevant 
SM parameters controlling quark-flavour dynamics (the quark masses 
and the angles of the Cabibbo-Kobayashi-Maskawa 
matrix~\cite{Cabibbo:1963yz,Kobayashi:1973fv}
have been determined with good accuracy.
More important, several suppressed observables
(such as $\Delta m_{B_d}$, $\Delta m_{B_s}$, $\cA^{_{\rm CP}}_{K \Psi}$,
$B\to X_s \gamma$, $\epsilon_K$, \ldots) potentially sensitive to 
physics beyond the SM have been measured with good accuracy,
showing no deviations from the SM.
The situation is somehow similar to the flavour-conserving 
electroweak precision observables after LEP: 
the SM works very well and genuine 
one-loop electroweak effects have been tested with 
relative accuracy in the $10\%$--$30\%$ range.
Similarly to the case of electroweak observables, also in the quark 
flavour sector non-standard effects can only appear as small 
corrections to the leading SM contribution.

Despite the success of the SM in electroweak and 
quark-flavour physics, we have also clear indications that 
this theory is not complete: the phenomenon of 
neutrino oscillations and the evidence for dark matter 
cannot be explained within the SM.  The 
SM is also affected by a serious theoretical problem 
because of the instability 
of the Higgs sector under quantum corrections.
We have not yet enough 
information to unambiguously determine how the SM Lagrangian 
should be modified; however, most realistic proposals point
toward the existence of new degrees of freedom in the TeV 
range, possibly accessible at the high-$p_T$ experiments
at the LHC. 

Assuming that the SM is not a complete theory, the precise tests 
of flavour dynamics performed so far imply a series of challenging 
constraints about the new theory:
if there are new degrees of freedom at the TeV scale,
present data tell us that they must posses a highly 
non-generic flavour structure. This structure is 
far from being established and its deeper investigation 
is the main goal of continuing high-precision flavour-physics 
in the LHC era. 
In these lectures we  focus on the interest 
and potential impact of future measurements in the $B$-meson 
system in this perspective.

The lectures are organized as follows: 
in the first lecture we briefly recall the main features of 
flavour physics within the SM. We also address in general 
terms the  so-called {\em flavour problem}, namely the challenge
to any SM extension posed by the success of the SM in flavour physics.
In the second lecture we analyse in some detail the 
SM predictions for some of the most interesting $B$ physics 
observables to be measured in the LHC era. 
In the last lecture we analyse flavour physics 
in various realistic beyond-the-SM scenarios, 
discussing how they can be tested by 
future experiments. 

The presentation of these lectures is somehow original, but all the 
material can be found, often with more details, in various 
set of lectures~\cite{Lectures} 
and review articles~\cite{Antonelli:2009ws,Buchalla:2008jp,Buras:2009if}
present in the recent literature.

\section{Flavour physics within the SM and the flavour problem}

\subsection{The  flavour sector of the SM}

The Standard Model (SM) Lagrangian can be divided into two main parts, 
the gauge and the Higgs (or symmetry breaking) sector. The gauge sector
is extremely simple and highly symmetric: it is completely specified by the 
local symmetry ${\mathcal G}^{\rm SM}_{\rm local} =SU(3)_{C}\times SU(2)_{L}\times U(1)_{Y}$
and by the fermion content,
\bea
\cL^{\rm SM}_{\rm gauge} &=& \sum_{i=1\ldots3}\ \sum_{\psi=Q^i_L \ldots E^i_R}
{\bar \psi} i D\sla \psi \no\\
&& -\frac{1}{4} \sum_{a=1\ldots8} G^a_{\mu\nu} G^a_{\mu\nu} -\frac{1}{4}
 \sum_{a=1\ldots3} W^a_{\mu\nu} W^a_{\mu\nu}  -\frac{1}{4} B_{\mu\nu} B_{\mu\nu}~.
\eea
The fermion content consist of five fields with different quantum numbers 
under the gauge group.\footnote{~The notation used to indicate each field
is $\psi(A,B)_Y$, where $A$ and $B$ denote the representation under the 
$SU(3)_{C}$ and $SU(2)_L$ groups, respectively,  and $Y$ is the $U(1)_Y$ charge.}
\be
\label{eq:SMfer}
Q^i_{L}(3,2)_{+1/6}~,\ \ U^i_{R}(3,1)_{+2/3}~,\ \
D^i_{R}(3,1)_{-1/3}~,\ \ L^i_{L}(1,2)_{-1/2}~,\ \ E^i_{R}(1,1)_{-1}~,
\ee
each of them appearing in three different replica or 
flavours ($i=1,2,3$).

This structure give rise to a large {\em global} flavour symmetry 
of $\cL^{\rm SM}_{\rm gauge}$.
Both the local and the global symmetries of $\cL^{\rm SM}_{\rm gauge}$
are broken with the introduction of a $SU(2)_L$ scalar doublet $\phi$,
or the Higgs field. The local symmetry is spontaneously 
broken by the vacuum expectation value of the Higgs field,
$\langle \phi \rangle  = v = (2\sqrt{2} G_F)^{-1/2} \approx 174$~GeV, while 
the global flavour symmetry is {\em explicitly broken} by 
the Yukawa interaction of $\phi$ with the fermion fields:
\be
\label{eq:SMY}
- {\cal L}^{\rm SM}_{\rm Yukawa}=Y_d^{ij} {\bar Q}^i_{L} \phi D^j_{R}
 +Y_u^{ij} {\bar Q}^i_{L} \tilde\phi U^j_{R} + Y_e^{ij} {\bar L}_{L}^i
\phi E_{R}^j + {\rm h.c.} \qquad  ( \tilde\phi=i\tau_2\phi^\dagger)~.  
\ee
The large global flavour symmetry of  $\cL^{\rm SM}_{\rm gauge}$, 
corresponding to the independent unitary rotations in flavour space 
of the five fermion fields in Eq.~(\ref{eq:SMfer}), is a $U(3)^5$ group. 
This can be decomposed as follows: 
\be 
{\mathcal G}_{\rm flavour} = U(3)^5 \times  
{\mathcal G}_{q} \times {\mathcal G}_{\ell}~, 
\label{eq:Gtot}
\ee
where 
\be
{\mathcal G}_{q} = {SU}(3)_{Q_L}\times {SU}(3)_{U_R} \times {SU}(3)_{D_R}, \qquad 
{\mathcal G}_{\ell} =  {SU}(3)_{L_L} \otimes {SU}(3)_{E_R}~.
\label{eq:Ggroups}
\ee
Three of the five $U(1)$ subgroups can be identified with the total barion and 
lepton number, which are not broken by $\cL_{\rm Yukawa}$, and the weak hypercharge, 
which is gauged and broken only spontaneously by $\langle \phi \rangle  
\not=0$. The subgroups controlling flavour-changing dynamics
and flavour non-universality  are the non-Abelian groups ${\mathcal G}_{q}$ 
and ${\mathcal G}_{\ell}$, which are explicitly broken by $Y_{d,u,e}$ not being 
proportional to the identity matrix. 

The diagonalization of each Yukawa coupling requires, in general, two 
independent unitary matrices, $V_L Y V^\dagger_R = {\rm diag}(y_1,y_2,y_3)$.
In the lepton sector the invariance of $\cL^{\rm SM}_{\rm gauge}$ 
under ${\mathcal G}_{\ell}$ allows us to freely choose the two matrices 
necessary to diagonalize  $Y_e$ without breaking gauge invariance, 
or without observable consequences. This is not the case in the quark 
sector, where we can freely choose only three of the four unitary matrices 
necessary to diagonalize both $Y_{d}$ and $Y_u$. Choosing the basis where 
$Y_{d}$ is diagonal (and eliminating the right-handed 
diagonalization matrix of $Y_u$) we can write 
\be
\label{eq:Ydbasis}
Y_d=\lambda_d~, \qquad  Y_u=V^\dagger\lambda_u~,
\ee
where 
\be
\label{eq:deflamd}
\lambda_d={\rm diag}(y_d,y_s,y_b)~, \ \ \
\lambda_u={\rm diag}(y_u,y_c,y_t)~, \qquad y_q = \frac{m_q}{v}~.
\ee
Alternatively we could choose a  gauge-invariant basis where 
$Y_d= V \lambda_d$ and $Y_u=\lambda_u$. Since the flavour symmetry  
do not allow the diagonalization from the left of both $Y_{d}$ and $Y_u$,
in both cases we are left with a non-trivial unitary mixing matrix, $V$, 
which is nothing but the Cabibbo-Kobayashi-Maskawa (CKM) 
mixing matrix~\cite{Cabibbo:1963yz,Kobayashi:1973fv}.

A generic unitary $3\times3$ [$N\times N$] complex unitary matrix depends 
on three [$N(N-1)/2$] real rotational angles and 
six [$N(N+1)/2$] complex phases. Having chosen a quark basis where 
the $Y_{d}$ and $Y_u$ have the form in (\ref{eq:Ydbasis})
leaves us with a  residual invariance under the flavour group
which allows us to eliminate five of the six complex phases in $V$ 
(the relative phases of the various quark fields).
As a result, the physical parameters in $V$ are four: three real angles and
one complex CP-violating phase. 
The full set of parameters controlling 
the breaking of the quark flavour symmetry in the SM is composed by the 
six quark masses in $\lambda_{u,d}$ and the four parameters of $V$.

For practical purposes it is often convenient to work in the mass eigenstate basis 
of both up- and and down-type quarks. This can be achieved rotating independently 
the up and down components of the quark doublet $Q_L$, or moving the CKM matrix 
from the Yukawa sector to the charged weak current in $\cL^{\rm SM}_{\rm gauge}$:
\be
\left. J_W^\mu \right|_{\rm quarks} = \bar u^i_L \gamma^\mu d^i_L \quad \stackrel{u,d~{\rm mass-basis}}{\longrightarrow} \quad
\bar u^i_L V_{ij} \gamma^\mu d^j_L ~.
\label{eq:Wcurrent}
\ee
However, it must be stressed that $V$ originates from the Yukawa sector (in particular 
by the miss-alignment of $Y_u$ and $Y_d$ in the ${SU}(3)_{Q_L}$ subgroup of ${\mathcal G}_q$): 
in absence of Yukawa  couplings we can always set $V_{ij}=\delta_{ij}$. 

To summarize, 
quark flavour physics within the SM is characterized by a large flavour symmetry, 
${\mathcal G}_{q}$, defined by the gauge sector, whose only breaking sources 
are the two Yukawa couplings $Y_{d}$ and $Y_{u}$. The CKM matrix arises by the 
miss-alignment of $Y_u$ and $Y_d$ in flavour space.

\subsection{Some properties of the CKM matrix}

The standard parametrization of the CKM matrix~\cite{Chau:1984fp}
in terms of three rotational angles ($\theta_{ij}$) 
and one complex phase ($\delta$) is 
\bea
V&=&\left(\begin{array}{ccc}
V_{ud}&V_{us}&V_{ub}\\
V_{cd}&V_{cs}&V_{cb}\\
V_{td}&V_{ts}&V_{tb}
\end{array}\right) \no \\
&=&
\left(\begin{array}{ccc}
c_{12}c_{13}&s_{12}c_{13}&s_{13}e^{-i\delta}\\ -s_{12}c_{23}
-c_{12}s_{23}s_{13}e^{i\delta}&c_{12}c_{23}-s_{12}s_{23}s_{13}e^{i\delta}&
s_{23}c_{13}\\ s_{12}s_{23}-c_{12}c_{23}s_{13}e^{i\delta}&-s_{23}c_{12}
-s_{12}c_{23}s_{13}e^{i\delta}&c_{23}c_{13}
\end{array}\right)~,
\label{eq:Chau}
\eea
where $c_{ij}=\cos\theta_{ij}$ and $s_{ij}=\sin\theta_{ij}$
($i,j=1,2,3$). 

The off-diagonal elements of the CKM matrix
show a strongly 
hierarchical pattern:  $|V_{us}|$ and $|V_{cd}|$ are close to $0.22$, the elements
$|V_{cb}|$ and $|V_{ts}|$ are of order $4\times 10^{-2}$ whereas $|V_{ub}|$ and
$|V_{td}|$ are of order $5\times 10^{-3}$. 
The Wolfenstein parametrization, namely the expansion of the CKM matrix 
elements in powers of the small parameter $\lambda \doteq |V_{us}| \approx 0.22$, is a
convenient way to exhibit this hierarchy in a more explicit way~\cite{Wolfenstein:1983yz}:
\begin{equation}
V=
\left(\begin{array}{ccc}
1-{\lambda^2\over 2}&\lambda&A\lambda^3(\varrho-i\eta)\\ -\lambda&
1-{\lambda^2\over 2}&A\lambda^2\\ A\lambda^3(1-\varrho-i\eta)&-A\lambda^2&
1\end{array}\right)
+{\cal{O}}(\lambda^4)~,
\label{eq:Wolfpar} 
\end{equation}
where $A$, $\varrho$, and $\eta$ are free parameters of order 1. 
Because of the smallness of $\lambda$ and the fact that for each element 
the expansion parameter is actually $\lambda^2$, this is a rapidly converging
expansion.

The Wolfenstein parametrization is certainly more transparent than
the standard parametrization. However, if one requires sufficient 
level of accuracy, the terms of ${\cal{O}}(\lambda^4)$ and 
${\cal{O}}(\lambda^5)$ have to be included in phenomenological applications.
This can be achieved in many different ways, according to the convention 
adopted. The simplest (and nowadays commonly adopted) choice is obtained 
{\it defining} the parameters $\{\lambda,A,\varrho,\eta\}$ in terms of 
the angles of the exact parametrization in Eq.~(\ref{eq:Chau}) as follows:
\begin{equation}
\label{eq:rhodef} 
\lambda\doteq s_{12}~,
\qquad
A \lambda^2\doteq s_{23}~,
\qquad
A \lambda^3 (\varrho-i \eta)\doteq s_{13} e^{-i\delta}~.
\end{equation}
The change of variables $\{ s_{ij}, \delta \} \to \{\lambda,A,\varrho,\eta\}$
in Eq.~(\ref{eq:Chau}) leads to an exact parametrization 
of the CKM matrix in terms of the Wolfenstein parameters.  
Expanding this expression up to ${\cal{O}}(\lambda^5)$ leads to
\begin{equation}
\label{eq:Buraspar} 
\left(\begin{array}{ccc}
1-\frac{1}{2}\lambda^2-\frac{1}{8}\lambda^4               &
\lambda+{\cal{O}}(\lambda^7)                                   & 
A \lambda^3 (\varrho-i \eta)                              \\
-\lambda+\frac{1}{2} A^2\lambda^5 [1-2 (\varrho+i \eta)]  &
1-\frac{1}{2}\lambda^2-\frac{1}{8}\lambda^4(1+4 A^2)     &
A\lambda^2+{\cal{O}}(\lambda^8)                                \\
A\lambda^3(1-\rhob-i\etab)                       &  
\!\!\!\!\!  -A\lambda^2+\frac{1}{2}A\lambda^4[1-2 (\varrho+i\eta)]   &
1-\frac{1}{2} A^2\lambda^4                           
\end{array}\right)
\end{equation}
where
\begin{equation}
\label{eq:rhobar}
\rhob = \varrho (1-\frac{\lambda^2}{2})+{\cal O}(\lambda^4)~,
\qquad
\etab=\eta (1-\frac{\lambda^2}{2})+{\cal O}(\lambda^4)~.
\end{equation}
The advantage of this generalization of the Wolfenstein parametrization
is the absence of relevant corrections to $V_{us}$, $V_{cd}$, $V_{ub}$ and 
$V_{cb}$, and a simple change in $V_{td}$, which facilitate the implementation 
of experimental constraints.

\begin{figure}[t]
\begin{center}
\includegraphics[width=7cm]{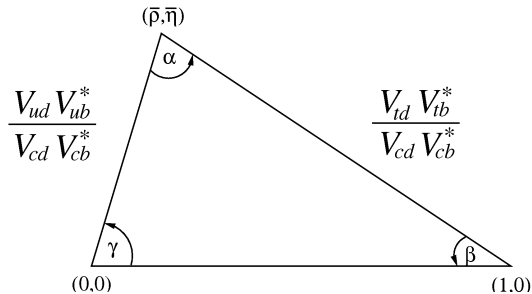}
\end{center}
\caption{\it The CKM unitarity triangle.}
\label{fig:1utriangle}
\end{figure}

The unitarity of the CKM matrix implies the following relations between its
elements:
\be
{\bf I)}\quad 
 \sum_{k=1\ldots 3} V_{ik}^* V_{ki}=1~,
\quad\qquad 
{\bf II)}\quad
\sum_{k=1\ldots 3} V_{ik}^* V_{kj\not=i}~.
\ee
These relations are a distinctive feature of the SM, where the CKM matrix is the only 
source of quark flavour mixing.  Their experimental verification is therefore a useful 
tool to set bounds, or possibly reveal, new sources of flavour symmetry breaking. 
Among the relations of type {\bf II}, the one obtained for $i=1$ and $j=3$, 
namely 
\be
V_{ud}^{}V_{ub}^* + V_{cd}^{}V_{cb}^* + V_{td}^{}V_{tb}^* =0 
\label{eq:UT}
\ee
\be
{\rm or} \qquad 
\frac{V_{ud}^{}V_{ub}^*}{V_{cd}^{}V_{cb}^*}  + \frac{V_{td}^{}V_{tb}^*}{V_{cd}^{}V_{cb}^*}  + 1 = 0
\qquad \leftrightarrow \qquad
 [\rhob+i \etab] + [(1-\rhob)-i\etab] + 1 =0~,
\no
\ee
is particularly interesting since it involves the sum of three terms all
of the same order in $\lambda$ and is usually represented as a unitarity triangle
in the complex  plane, as shown in Fig.~\ref{fig:1utriangle}.
It is worth to stress that Eq.~(\ref{eq:UT}) is invariant under any 
phase transformation of the quark fields. Under such transformations
the triangle in Fig.~\ref{fig:1utriangle} is rotated in the complex plane,
but its angles and the sides remain unchanged.
Both angles and  sides of the unitary triangle are indeed observable quantities
which can be measured in suitable experiments.  

\subsection{Present status of CKM fits}
\label{sect:CKMfits} 

\begin{figure}[t]
  \centering
  {\includegraphics[width=0.6\textwidth]{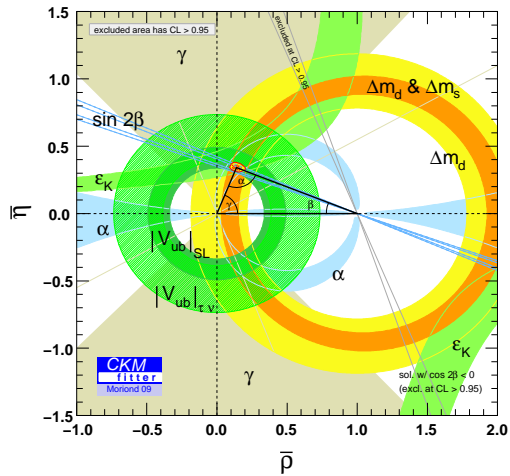}}
  \caption{Allowed region in the $\rhob,\etab$ plane, from~\cite{ckmfitter}. 
  Superimposed are
  the individual constraints from charmless semileptonic $B$ decays
  ($|V_{ub}|$), mass differences in the $B_d$ ($\Delta m_d$)
  and $B_s$ ($\Delta m_s$) systems, CP violation 
  in the neutral kaon ($\varepsilon_K$) and in the $B_d$ systems ($\sin2\beta$),
  the combined constrains on $\alpha$ and $\gamma$ from various $B$ decays. }
  \label{fig:UT}
\end{figure}

The values of $|V_{us}|$ and $|V_{cb}|$, or $\lambda$ and $A$ in the parametrization 
(\ref{eq:Buraspar}), are determined with good accuracy from  $K\to\pi\ell\nu$ and 
$B\to X_c \ell\nu$ decays, respectively. According to the recent analysis
in Ref.~\cite{Antonelli:2009ws}, their numerical values are
\be
\lambda=0.2259 \pm0.0009~,\qquad A=0.802\pm0.015~.
\ee
Using these results, all the other constraints on the elements of 
the CKM matrix can be expressed as constraints on $\rhob$ and $\etab$
(or constraints on the CKM unitarity triangle in Fig.~\ref{fig:1utriangle}).
The list of the most sensitive observables used to determine  
$\rhob$ and $\etab$ in the SM includes:
\begin{itemize}
\item The rates of inclusive and exclusive charmless semileptonic $B$
   decays, which depend on $|V_{ub}|$ and provide a constraint on $\rhob^2+\etab^2$.
\item The time-dependent CP asymmetry in $B\to\psi K_S$ decays 
   ($\cA^{_{\rm CP}}_{K \Psi}$), which depends on the phase of the 
   $B_d$--$\bar B_d$ mixing amplitude relative to the decay amplitude 
   (see Sect.~\ref{sect:BBmix}).
   Within the SM this translates into a constraint on  $\sin2\beta$.
\item The rates of various $B\to DK$ decays, which provide constraints on the 
  angle  $\gamma$  (see Sect.~\ref{sect:gamma}).
\item The rates of various $B\to\pi\pi,\rho\pi,\rho\rho$ decays, which 
   constrain $\alpha=\pi-\beta-\gamma$.
\item The ratio between the mass splittings in the neutral $B$ and
  $B_s$ systems, which depends on $|V_{td}/V_{ts}|^2 \propto [(1-\rhob)^2+\etab^2]$.
\item The indirect CP violating parameter of the kaon system 
  ($\epsilon_K$), which determines and hyperbola in the $\rhob$ and $\etab$
  plane (see  Ref.~\cite{Antonelli:2009ws} for more details). 
\end{itemize}
The resulting constraints are shown in Fig.~\ref{fig:UT}.
As can be seen, they are all consistent with a unique value of 
$\rhob$ and $\etab$~\cite{Antonelli:2009ws}: 
\be
\rho=+0.158 \pm 0.021~, \qquad \eta=+0.343\pm0.013~.
\ee

The consistency of different constraints 
on the CKM unitarity triangle is a powerful consistency test 
of the SM in describing flavour-changing phenomena.
From the plot in Fig.~\ref{fig:UT} it is quite clear, 
at least in a qualitative way, that there is little room 
for non-SM contributions in flavour changing transitions. 
A more quantitative evaluation of this 
statement is presented in the next section.

\subsection{The SM as an effective theory} 
\label{sect:effth}

As anticipated in the introduction, despite the impressive phenomenological 
success of the SM in flavour and electroweak physics, there are various 
convincing arguments which motivate us to consider this model only as the 
low-energy limit of a more complete theory.

Assuming that the new degrees of freedom which complete the theory 
are heavier than the SM particles,
we can integrate them out and describe physics beyond the SM in full
generality by means of an {\em effective theory} approach.
The SM Lagrangian becomes the renormalizable part of a more 
general local Lagrangian which includes an infinite tower of 
operators with dimension $d>4$, constructed in terms of SM fields
and suppressed by inverse powers of an effective scale 
$\Lambda$. These operators are the residual effect of 
having integrated out the new heavy degrees of freedom, whose
mass scale is parametrized by the effective scale $\Lambda > m_W$. 

As we will discuss in more detail in Sect.~\ref{sect:Heff}, 
integrating out heavy degrees of freedom is a procedure
often adopted also within the SM: it allows us to 
simplify the evaluation of amplitudes which involve 
different energy scales. This approach is indeed
a generalization of the Fermi theory of weak interactions,
where the dimension-six four-fermion operators describing 
weak decays are the results of having integrated out 
the $W$ field. The only difference when applying this 
procedure to physics beyond the SM is that in this case,
as also in the original work by Fermi, we don't know the 
nature of the degrees of freedom we are integrating out. 
This imply  we are not able to determine a priori the 
values of the effective couplings of the higher-dimensional 
operators. The advantage of this approach is that it 
allows us to analyse all realistic extensions of the SM 
in terms of a limited number of parameters 
(the coefficients of the higher-dimensional operators). 
The drawback is the impossibility 
to establish correlations 
of New Physics (NP) effects at low and high energies.

Assuming for simplicity that there is a single elementary 
Higgs field, responsible for the $SU(2)_L \times U(1)_Y \to U(1)_Q$
spontaneous breaking, the Lagrangian of the SM considered 
as an effective theory can be written as follows
\be
\cL_{\rm eff} = \cL^{\rm SM}_{\rm gauge} + \cL^{\rm SM}_{\rm Higgs} + 
\cL^{\rm SM}_{\rm Yukawa}
+\Delta \cL_{d > 4}~,
\ee
where $\Delta \cL_{d > 4}$ denotes the series of higher-dimensional 
operators invariant under the SM gauge group:
\be
\Delta \cL_{d > 4}  = ~\sum_{d > 4} ~\sum_{n=1}^{N_d} ~
\frac{c^{(d)}_n}{\Lambda^{d-4}} \cO^{(d)}_n ({\rm SM~fields}). 
\label{eq:DL_eff}
\ee
If NP appears at the TeV scale, as we expect from the stabilization 
of the mechanism of electroweak symmetry breaking, the scale 
$\Lambda$ cannot exceed a few TeV.  Moreover, if the underlying 
theory is natural  (no fine-tuning in the coupling constants), we 
expect $c^{(d)}_i=O(1)$ for all the operators which 
are not forbidden (or suppressed) by symmetry arguments. 
The observation that this expectation is {\em not} 
fulfilled by several dimension-six operators contributing 
to flavour-changing processes is often denoted 
as the {\em flavour problem}.

If the SM Lagrangian were invariant under some flavour symmetry, 
this problem could easily be circumvented. 
For instance in the case of  barion- or lepton-number violating 
processes, which are exact symmetries of the SM Lagrangian,
we can avoid the tight experimental bounds promoting 
$B$ and $L$ to be exact symmetries of the new dynamics 
at the TeV scale. The peculiar aspects of flavour 
physics is that there is no exact flavour symmetry 
in the low-energy theory. In this case it is 
not sufficient to invoke a flavour symmetry
for the underlying dynamics. We also need to specify how this 
symmetry is broken in order to describe the observed low-energy 
spectrum and, at the same time, be in agreement 
with the precise experimental tests
of flavour-changing processes. 

\subsubsection{Bounds on the scale of New Physics from $\Delta F=2$
processes}
\label{sect:DF2bounds}

The best way to quantify the flavour problem is obtained 
by looking at consistency of the tree- and loop-mediated 
constraints on the CKM matrix 
discussed in Sect.~\ref{sect:CKMfits}. 

In first approximation we can assume that NP effects are
negligible in processes which are dominated by tree-level 
amplitudes. Following this assumption, the values of 
$|V_{us}|$, $|V_{cb}|$, and $|V_{ub}|$, as well as the 
constraints on $\alpha$ and $\gamma$ 
are essentially NP free. As can be seen in Fig.~\ref{fig:UT},
this implies we can determine completely the CKM matrix assuming generic 
NP effects in loop-mediated amplitudes.
We can then use the measurements of observables which are
loop-mediated within the SM to bound the couplings of 
effective NP operators in Eq.~(\ref{eq:DL_eff})
which contribute to these observables at the tree level.

\begin{figure}[t]
  \centering
  {\includegraphics[width=0.6\textwidth]{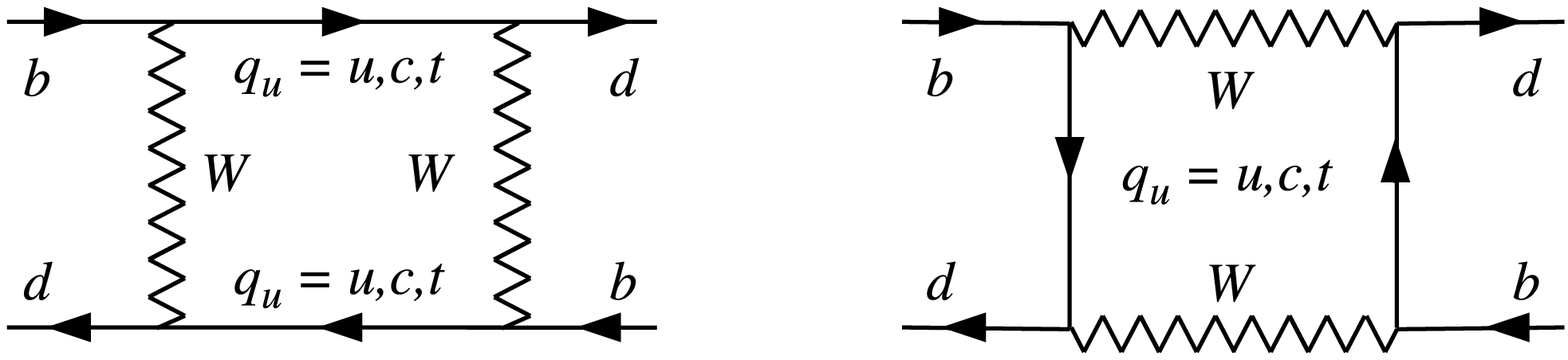}}
 \caption{Box diagrams contributing to $B_d$-$\bar B_d$ mixing in the unitary gauge.}
  \label{fig:BBmix}
\end{figure}

The loop-mediated constraints shown in Fig.~\ref{fig:UT}
are those from the mixing of $B_d$, $B_s$, and $K^0$
with the corresponding anti-particles
(generically denoted as $\Delta F=2$ amplitudes).
Within the SM, these amplitudes are generated by 
box amplitudes of the type in Fig.~\ref{fig:BBmix}
(and similarly for $B_s$, and $K^0$) and are affected 
by small hadronic uncertainties (with the exception of 
$\Delta m_K$). 
We will come back to the evaluation of these amplitudes 
in more detail in Sect.~\ref{sect:BBmix}. For the moment 
it is sufficient to notice that the leading contribution 
is obtained with the top-quark running 
inside the loop, giving rise to the highly suppressed 
result
\be
 \cM_{\Delta F=2}^{\rm SM} \approx  \frac{ G_F^2 m_t^2 }{16 \pi^2} ~ V_{3i}^* V_{3j} ~
    \langle \bar  M |  (\bar d_L^i \gamma^\mu d_L^j )^2  
| M \rangle \times  F\left(\frac{m_t^2}{m_W^2}\right)
\qquad [M = K^0, B_d, B_s]~,
\ee
where $F$ is a loop function of order one 
($i,j$ denote the flavour indexes of the meson
valence quarks).

Magnitude and phase of all these
mixing amplitudes have been determined 
with good accuracy from experiments
with the exception of the
CP-violating phase in $B_s$--$\bar B_s$ mixing.
As shown in  Fig.~\ref{fig:UT}, 
in all cases where the experimental information is 
precise, the magnitude of the new-physics amplitude
cannot exceed, in size, the SM contribution.

To translate this information into bounds on the scale 
of new physics, let's consider the following set of 
$\Delta F=2$ dimensions-six operators
\be
\cO_{\Delta F=2}^{ij} = (\bar Q_L^i \gamma^\mu Q_L^j )^2 ~,
\qquad Q^i_L =\left(\ba{c} u^i_L \\ d^i_L \ea \right)~,
\label{eq:dfops}
\ee
where $i,j$ are flavour indexes in the 
basis defined by Eq.~(\ref{eq:Ydbasis}). 
These operators contribute at the tree-level to 
the meson-antimeson mixing amplitudes.
Denoting $c_{ij}$ the couplings of the non-standard 
operators in (\ref{eq:dfops}), the condition 
$| \cM_{\Delta F=2}^{\rm NP}| <  | \cM_{\Delta F=2}^{\rm SM} |$
implies\footnote{~A more refined analysis, with complete 
statistical treatment and separate bounds form the real 
and the imaginary parts of the various amplitudes,
leading to slightly more stringent bounds, 
can be found in~\cite{Bona:2007vi}.}
\bea
\Lambda < \frac{ 3.4~{\rm TeV} }{| V_{3i}^* V_{3j}|/|c_{ij}|^{1/2}  }
<  \left\{ \ba{l}  
9\times 10^3~{\rm TeV} \times |c_{21}|^{1/2} \qquad {\rm from} \quad 
K^0-\bar K^0
 \\
4\times 10^2~{\rm TeV} \times |c_{31}|^{1/2} \qquad {\rm from}  \quad
B_d-\bar B_d
 \\
7\times 10^1~{\rm TeV} \times |c_{32}|^{1/2} \qquad {\rm from}  \quad
B_s-\bar B_s \ea
\right. 
\label{eq:boundsDF2}
\eea 
The main messages of these bounds are the following:
\begin{itemize}
\item New physics models with a generic flavour 
structure ($c_{ij}$ of order 1) at the TeV scale are
ruled out. If we want 
to keep $\Lambda$ in the TeV range, physics beyond the SM
must have a highly non-generic flavour structure.  
\item In the specific
case of the $\Delta F=2$ operators in  (\ref{eq:dfops}),
in order to  keep $\Lambda$ in the TeV range,
we must find a symmetry argument such that  
$|c_{ij}| \lsim  |V_{3i}^* V_{3j}|^2$.
\end{itemize}

The strong constraining power of $\Delta F=2$ observables 
is a consequence of their strong suppression within the SM.
They are suppressed not only 
by the typical $1/(4\pi)^2$ factor of loop amplitudes, 
but also by the GIM mechanism~\cite{Glashow:1970gm}
 and by the hierarchy of the 
CKM matrix ($|V_{3i}|\ll 1$, for $i\not =3$). 
Reproducing a similar structure beyond the SM is a highly
non-trivial task. As we will discuss in the last lecture,
only in a few cases this can be implemented in a natural way.

To conclude, we stress that the good agreement of SM 
and experiments for $B_d$ and $K^0$ mixing does not imply that 
further studies of flavour physics are not interesting. 
On the one hand, even for $|c_{ij}| \approx  |V_{3i}^* V_{3j}|$,
which can be considered the most pessimistic case, as we will 
discuss in Sect.~\ref{sect:MFV},
we are presently constraining physics at the TeV scale. 
Therefore improving these bounds, if possible, would be
extremely valuable. One the other hand, as we will discuss in the next lecture, 
there are various interesting observables which have not been deeply 
investigated yet, whose study could reveal additional
key features about the flavour structure 
of physics beyond the SM.

\section{$B$-physics phenomenology: 
            mixing, CP violation, and rare decays}

As we have seen in the previous lecture, the exploration 
of the mechanism of quark-flavour mixing is entered in a
new era. The precise measurements 
of mixing-induced CP violation and tree-level allowed
semileptonic transition have provided an important
consistency check of the SM, and a precise determination 
of the Cabibbo-Kobayashi-Maskawa  matrix. 
The next goal is to understand if there is still 
room for new physics or, more precisely,
if there is still room for new sources of flavour symmetry 
breaking close to the electroweak scale.
From this perspective, a few rare $B$ decays
mediated by flavour-changing neutral-current (FCNC) 
amplitudes, and also the CP violating phase of 
$B_s$ mixing, represent a fundamental tool. 

Beside the experimental sensitivity, 
the conditions which allow us to perform significant NP searches 
in rare decays can be summarized as follows:
i) decay amplitude dominated by electroweak dynamics,
and thus enhanced sensitivity to non-standard contributions;
ii) small theoretical error within the SM, or good control 
of both perturbative and non-perturbative corrections.
In the next section we analyze how and in which cases 
these two conditions can be achieved.

\subsection{Theoretical tools: low-energy effective Lagrangians}
\label{sect:Heff}

The decays of $B$ mesons are processes which involve at least two different 
energy scales: the electroweak scale, characterized by the $W$ boson mass, 
which determines the flavor-changing transition at the quark level, 
and the scale of strong interactions $\Lambda_\mathrm{QCD}$,
related to the hadron formation. The presence of these two widely separated 
scales makes the calculation of the decay amplitudes starting from the 
full SM Lagrangian quite complicated: large logarithms of the type 
log($m_W/\Lambda_\mathrm{QCD}$) may appear, leading to a breakdown of 
ordinary perturbation theory. 

This problem can be substantially simplified by integrating out the 
heavy SM fields ($W$ and $Z$ bosons, as well as the top quark)
at the electroweak scale, and constructing an appropriate low-energy 
effective theory where only the light SM fields appear.
The weak effective Lagrangians thus obtained 
contains local operators of dimension six (and higher), written 
in terms of light SM fermions, photon and gluon fields, suppressed 
by inverse powers of the $W$ mass. 

To be concrete, let's consider the example of charged-current 
semileptonic weak interactions. The basic building 
block in the full SM Lagrangian is
\be
\cL^{\rm full~SM}_{W} = \frac{g}{\sqrt{2}} J_{W}^\mu(x) W_\mu^+(x) + {\rm h.c.}~,
\ee
where 
\be
J_W^\mu(x)= V_{i j}~\bar u_L^i(x)\gamma^\mu d_L^j(x)
+\bar e^j_L(x) \gamma^\mu \nu_L^j(x) 
\ee
is the weak charged current already introduced in Eq.~(\ref{eq:Wcurrent}).
Integrating out the $W$ field at the tree level we contract
two vertexes of this type generating the  non-local transition amplitude
\be
i \cT = - i \frac{g^2}{2} \int d^4x D_{\mu \nu} \left( x, m_W \right) 
T \left[ J_W^\mu(x),J_W^{\nu\dagger}(0) \right]~,
\ee
which involves only light fields. Here $D_{\mu \nu} \left( x, m_W \right)$ is 
the $W$ propagator in coordinate space: expanding it in inverse powers of $m_W$,
\begin{equation}
D_{\mu \nu} \left( x, m_W \right)=\int \frac{d^4 q}{(2\pi)^4} e^{-i q\cdot x} \frac{-i g_{\mu\nu} + \cO(q_\mu,q_\nu) }
{q^2-m_W^2+i\varepsilon}=\delta(x) \frac{ i g_{\mu\nu}}{m_W^2}+\dots\,,
\label{eq:MWexp}
\end{equation}
the leading contribution to $\cT$  can be interpreted as  
the tree-level contribution of the following effective local Lagrangian 
\be
\cL^{\rm (0)}_{\rm eff} = - \frac{4 G_F}{\sqrt{2}}  g_{\mu\nu} 
J_W^\mu(x) J_W^{\nu\dagger}(x)~,
\label{eq:LFermi0}
\ee
where  $G_F/\sqrt{2}=g^2/(8 m_W^2)$ is the Fermi coupling.
If we select in the product of the two currents
one quark and one leptonic current, 
\be
\cL^{\rm semi-lept}_{\rm eff} = - \frac{4 G_F}{\sqrt{2}}~V_{i j}~\bar u_L^i(x)\gamma^\mu d_L^j(x)
~\bar \nu_L(x) \gamma_\mu e_L(x) + {\rm h.c.}~,
\label{eq:LFermi}
\ee
we obtain an effective Lagrangian 
which provides an excellent description of semileptonic weak decays.
The neglected terms in the expansion (\ref{eq:MWexp}) correspond
to corrections of $\cO(m_B^2/m^2_W)$ to the decay amplitudes. In principle,
these corrections could be taken into account by adding appropriate 
dimension-eight operators in the effective Lagrangian. However, in most 
cases they are safely negligible.

The case of charged semileptonic decays is particularly simple since we can 
ignore QCD effects: the operator (\ref{eq:LFermi}) is not renormalized 
by strong interactions. The situation is slightly more 
complicated in the case of non-leptonic or flavour-changing neutral-current
processes, where QCD corrections and higher-order weak interactions 
cannot be neglected, but the basic strategy is the same. 
First of all we need to identify a complete basis of local operators, 
that includes also those generated beyond the tree level. In general,
given a fixed order in the  $1/m^2_W$ expansion of the amplitudes,
we need to consider all operators of corresponding dimension 
(e.g.~dimension six at the first order in the  $1/m^2_W$ expansion)
compatible with the symmetries of the system. Then we must introduce 
an artificial scale in the problem, the renormalization scale $\mu$,
which is needed to regularize QCD (or QED) corrections 
in the effective theory.

The effective Lagrangian for generic $\Delta F=1$ processes assumes the form
\be
\cL_{\Delta F=1} =  - 4\frac{G_F}{\sqrt{2}}  \sum_i C_i (\mu)  Q_i 
\label{eq:effH} 
\ee
where the sum runs over the complete basis of operators.
As explicitly indicated, the effective couplings $C_i(\mu)$
(known as Wilson coefficients) depend, in general,
on the renormalization scale. The dependence from this
scale cancels when evaluating the matrix elements of the effective 
Lagrangian for physical processes, that we 
can generically indicate as 
\be
\cM (i\to f) =  - 4\frac{G_F}{\sqrt{2}}  \sum_i C_i (\mu) \langle f |  Q_i(\mu) | i \rangle~.
\ee
The independence of $\cM$ from $\mu$ holds for any initial and final state,
including partonic states at high energies.
This implies that the $C_i (\mu)$ obey a series of renormalization group 
equations (RGE), whose structure is completely determined 
by the anomalous dimensions of the effective operators. 
These equations can be solved using standard RG techniques,
allowing the resummation of all large logs of the type 
 $\alpha_s(\mu)^{n+m}\log(m_W/\mu)^n$ to all orders in $n$
(working at order $m+1$ in perturbation theory).
The scale $\mu$ acts as a separator of short- and long-distance virtual corrections:
short-distance effects are included in the  $C_i(\mu)$,
whereas long-distance effects are left as explicit degrees
of freedom in the effective theory.\footnote{~This statement would 
be correct if the theory were regularized using a dimensional 
cut-off. It is not fully correct if $\mu$ is the scale 
appearing in the (often adopted) dimensional-regularization 
+ minimal-subtraction (MS) renormalization scheme.}

In practice, the problem reduces to the following three 
well-defined and independent steps:
\begin{enumerate}
\item[1.] the evaluation of the {\em initial conditions} of the 
$C_i (\mu)$ at the electroweak scale $(\mu \approx m_W)$;
\item[2.] the evaluation of the anomalous dimension of the 
effective operators, and the corresponding {\em RGE evolution} 
of the $C_i (\mu)$ from the electroweak scale down 
to the energy scale of the physical process ($\mu \approx m_B$);
\item[3.] the evaluation of the {\em matrix elements} of the 
effective Lagrangian for the physical hadronic processes
(which involve energy scales from $m_B$ down to $\Lambda_{QCD}$).
\end{enumerate}

The first step is the one where physics beyond the SM may appear:
if we assume NP is heavy, it may modify the initial 
conditions of the Wilson coefficients at the high scale, 
while it cannot affect the following two steps.
While the RGE evolution and the hadronic matrix elements are not 
directly related to NP, they may influence the sensitivity to NP 
of physical observables. In particular, the evaluation of 
hadronic matrix elements is potentially affected by non-perturbative 
QCD effects: these are often a large source of theoretical uncertainty 
which can obscure NP effects. RGE effects do not induce sizable uncertainties
since they can be fully handled within perturbative QCD;
however, the sizable logs generated by the RGE running may {\em dilute}
the interesting short-distance information encoded in the 
value of the Wilson coefficients at the high scale.
As we will discuss in the following, only in specific 
observables these two effects are small and under good theoretical control.

A deeper discussion about the construction of low-energy effective Lagrangians, 
with a detailed discussions of the first two steps mentioned above, can be found in 
Ref.~\cite{Buchalla:1995vs}.

\subsubsection{Effective operators for rare decays and $\Delta F=2$ amplitudes}
\label{sect:effweak}

Let's give a closer look to processes where the underlying parton process
is $b\to s + \bar q q$. In this case the relevant effective Lagrangian 
can be written as
\be
\cL_{b \to s}^{\rm non-lept} = - 4 \frac{G_F}{\sqrt{2}} \left(
\sum_{q=u,c} \lambda_q^s  \sum_{i=1,2} C_i(\mu) Q^q_i(\mu) 
-\lambda_t^s \sum_{i=3}^{10} C_i(\mu) Q_i(\mu)\right)~,
\label{eq:hdb1}
\ee
where $\lambda^{s}_{q}=V^*_{qb} V_{qs}$, and the operator basis is 
\be
\begin{array}{ll}
Q^q_{1} = {\bar b}_L^\alpha\gamma^\mu q_L^\alpha\, {\bar q}_L^\beta\gamma_\mu s_L^\beta~, &
Q^q_{2} = {\bar b}_L^\alpha\gamma^\mu q_L^\beta\, {\bar q}_L^\beta\gamma_\mu s_L^\alpha~, \\
Q_{3} = {\bar b}_L^\alpha \gamma^\mu s_L^\alpha\, \sum_q {\bar q}_L^\beta\gamma_\mu q_L^\beta~, &
Q_{4} = {\bar b}_L^\alpha \gamma^\mu s_L^\beta\, \sum_q {\bar q}_L^\beta\gamma_\mu q_L^\alpha~, \\
Q_{5} = {\bar b}_L^\alpha \gamma^\mu s_L^\alpha\, \sum_q {\bar q}_R^\beta\gamma_\mu q_R^\beta~,  
\qquad\qquad &
Q_{6} = {\bar b}_L^\alpha \gamma^\mu s_L^\beta\, \sum_q {\bar q}_R^\beta\gamma_\mu q_R^\alpha~, \\
Q_{7} = \frac{3}{2}{\bar b}_L^\alpha \gamma^\mu s_L^\alpha\, \sum_q e_q {\bar q}_R^\beta\gamma_\mu q_R^\beta~, &
Q_{8} = \frac{3}{2}{\bar b}_L^\alpha \gamma^\mu s_L^\beta\, \sum_q e_q {\bar q}_R^\beta\gamma_\mu q_R^\alpha~, \\
Q_{9} = \frac{3}{2}{\bar b}_L^\alpha \gamma^\mu s_L^\alpha\, \sum_q e_q {\bar q}_L^\beta\gamma_\mu q_L^\beta~, &
Q_{10} =  \frac{3}{2}{\bar b}_L^\alpha \gamma^\mu s_L^\beta\, \sum_q e_q {\bar q}_L^\beta\gamma_\mu q_L^\alpha~,\\
\end{array}
\label{eq:basis}
\ee
with $\{\alpha,\beta\}$ and $e_q$ denoting color indexes
the electric charge of the quark $q$, respectively. 

Out of these operators, only $Q^c_1$ and $Q^u_1$ are generated at the tree-level
by the $W$ exchange. Indeed, comparing with the tree-level structure in~(\ref{eq:LFermi0}), 
we find 
\be 
C^{u,c}_1(m_W) = 1 + \cO(\alpha_s,\alpha)~, \qquad 
 C^{u,c}_{2-10}(m_W) = 0 + \cO(\alpha_s,\alpha)~.
\ee
However, after including RGE effects and running down to $\mu\approx m_b$, 
both $C^{u,c}_1$ and $C^{u,c}_2$ become $\cO(1)$, while 
$C_{3-6}$ become $\cO(\alpha_s(m_b))$. In all these cases there 
is little hope to identify NP effects: the leading initial condition 
is the tree-level $W$ exchange, which is hardly modified by NP.
In principle, the coefficients of the electroweak penguin operators, 
$Q_7$--$Q_{10}$, are more interesting: their initial conditions are 
related to electroweak penguin and box diagrams. However, it is 
hard to distinguish their contribution from those of the other 
four-quark operators in non-leptonic processes. Moreover, 
also for $C_{7-10}$ the relative contribution from long-distance 
physics (running down from  $m_W$ to $m_b$) 
is sizable and dilute the interesting short-distance 
information.

\medskip
 
For $b\to s$ transitions with a photon or a lepton pair in the final state, additional
dimension-six operators must be included in the basis, 
\be
\cL_{b \to s}^{\rm rare} = \cL_{b \to s}^{\rm non-lept} +  4 \frac{G_F}{\sqrt{2}} \lambda_t^s
\left(C_{7\gamma} Q_{7\gamma} + C_{8 g} Q_{8 g} +  C_{9 V} Q_{9 V} + 
 C_{10 A} Q_{10 A}\right)~,
\ee
where 
\begin{eqnarray}
&& Q_{7\gamma} = \frac{e}{16\pi^2} m_b {\bar b}_R^\alpha\sigma^{\mu\nu} F_{\mu\nu} s_L^\alpha~, \qquad\
Q_{8g} = \frac{g_s}{16\pi^2} m_b {\bar b}_R^\alpha\sigma^{\mu\nu} G_{\mu\nu}^A T^A s_L^\alpha~,   \nonumber\\
&& Q_{9V} = \frac{1}{2}{\bar b}_L^\alpha \gamma^\mu s_L^\alpha\, \bar l \gamma_\mu l~, \qquad\qquad\quad 
Q_{10A} = \frac{1}{2}{\bar b}_L^\alpha \gamma^\mu s_L^\alpha\, \bar l \gamma_\mu\gamma_5 l~,
\label{eq:radbasis}
\end{eqnarray}
and  $G^A_{\mu\nu}$ ($F_{\mu\nu}$) is the gluon (photon) field strength tensor. 
The initial conditions of these 
operators are particularly sensitive to NP: within the SM
they are generated by one-loop penguin and box diagrams 
dominated by the top-quark exchange. The most theoretically 
clean is $C_{10A}$, which do not mix with any of the 
four-quark operators listed above and which has a vanishing 
anomalous dimension:
\be C^{\rm SM}_{10A}(m_W) = \frac{g^2}{8\pi^2} \frac{x_t}{8}\left[ \frac{4- x_t}{1-x_t}
  +\frac{3x_t}{(1-x_t)^2}\;\ln x_t\right]~, \qquad x_t = \frac{m_t^2}{m_W^2}~.
\label{eq:Y0}
\ee  
NP effects at the TeV scale could easily modify this result, and this deviation 
would directly show up in low-energy observables sensitive to $C_{10A}$, such as 
$\cA_{\rm FB}(B\to K^*\ell^+\ell^-)$ and $\cB(B\to \ell^+\ell^-)$ (see Sect.~\ref{sect:BKll}  and \ref{sect:Bll}). We finally note that while the 
operators in Eqs.~(\ref{eq:basis}) and (\ref{eq:radbasis}) form a complete basis within the SM,
this is not necessarily the case beyond the SM. In particular, within 
specific scenarios also right-handed current operators
(e.g.~those obtained from  (\ref{eq:radbasis}) for $q_{L(R)} \to q_{R(L)}$)
may appear.

\medskip

The $\Delta F=2$ effective weak Lagrangians are simpler than the $\Delta F=1$ ones: 
the SM operator basis includes one operator only. The Lagrangian relevant for 
$B^0_d$--$\bar B^0_d$ and $B^0_s$--$\bar B^0_s$ mixing is conventionally 
written as ($q=\{d,s\}$):
\begin{equation}
{\cal L}^{\rm SM}_{\Delta B=2} = \frac{G_F^2}{4\pi^2}m_W^2 (V_{tb}^*V_{tq})^2~\eta_B(\mu)~S_0(x_t)~
(\bar b_L \gamma_\mu q_L\, \bar b_L \gamma^\mu q_L)~,
\label{eq:db2}
\end{equation}
where the initial condition of the Wilson coefficient is the loop function $S_0(x_t)$, 
corresponding to the box diagrams in Fig.~\ref{fig:BBmix}. The effect of QCD
correction is only a multiplicative correction factor, $\eta_B(\mu)$, which
can be computed with high accuracy and turns out to be of order one.
The explicit expression of the loop function, dominated by the top-quark exchange, is 
\be
S_0(x_t) = \frac{4x_t-11x_t^2+x_t^3}{4(1-x_t)^2} -\frac{3x_t^3 \ln x_t}{2(1-x_t)^3}~.
\label{eq:S0}
\ee  

\subsubsection{The gaugeless limit of FCNC and $\Delta F=2$ amplitudes}

\begin{figure}[t]
\vskip 0.5 cm
  \centering
  {\includegraphics[width=0.6\textwidth]{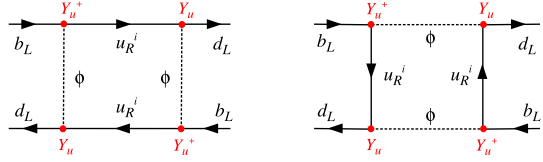}}
\vskip 0.5 cm
 \caption{One-loop contributions $\Delta F=2$ amplitudes in the gaugeless limit.}
  \label{fig:Gaugeless}
\end{figure}

An interesting aspect which is common to the electroweak loop
functions in Eqs.~(\ref{eq:Y0}) and (\ref{eq:S0}) is the fact they 
diverge in the limit $m_t/m_W \to \infty$. This behavior is 
apparently strange: it contradicts the expectation that 
contributions of heavy particles at low energy  
decouple in the limit where their masses  increase. 
The origin of this effect can be understand by noting that the 
leading contributions to both amplitudes are generated only by the Yukawa
interaction. These contributions can be better isolated in the {\em gaugeless} 
limit of the SM, i.e.~if we send to zero the gauge couplings.
In this limit $m_W\to 0$ and the derivation of the effective 
Lagrangian discussed in Sect.~\ref{sect:Heff} does not make sense.
However, the leading contributions to the effective 
Lagrangians for $\Delta F=2$ and rare decays are unaffected.
Indeed, the leading contributions to these processes are 
generated by Yukawa interactions of the type in Fig.~\ref{fig:Gaugeless}, 
where the scalar fields are the Goldstone-bosons components of the 
Higgs field (which are not eaten up by the $W$ in the limit $g\to0$).
Since the top is still heavy, we can integrate it out, obtaining
the following result for ${\cal L}_{\Delta B=2}$:
\be
\left. {\cal L}^{\rm SM}_{\Delta B=2} \right|_{g_i \to 0} ~=~ \frac{G_F^2 m_t^2}{16 \pi^2} 
 (V_{tb}^*V_{tq})^2 (\bar b_L \gamma_\mu q_L)^2 ~ = ~
\frac{[(Y_u Y^*_u)_{bq}]^2}{128 \pi^2 m_t^2}  (\bar b_L \gamma_\mu q_L)^2~.
\label{eq:gaugeless}
\ee
Taking into account that $S_0(x)\to x/4$
for  $x \to \infty$, it is easy to verify that this result is equivalent 
to the one in Eq.~(\ref{eq:S0}) in the large $m_t$ limit. 
A similar structure holds for the $\Delta F=1$ amplitude 
contributing to the axial operator $Q_{10A}$.

The last expression in Eq.~(\ref{eq:gaugeless}), which holds in the limit where we
neglect the charm Yukawa coupling, shows that the decoupling of the amplitude 
with the mass of the top is compensated by four powers of the 
top Yukawa coupling at the numerator. The divergence for  $m_t\to \infty$
can thus be understood as the divergence of one of the fundamental 
couplings of the theory.  Note also that in the gaugeless limit there 
is no GIM mechanism: the contributions of the various up-type quarks 
inside the loops do not cancel each other: they are directly weighted 
by the corresponding Yukawa couplings, and this is why the top-quark 
contribution is the dominant one. 

This exercise illustrates the key 
role of the Yukawa coupling in determining the main properties
flavour physics within the SM, as advertised in the first lecture.
It also illustrates the interplay of flavour and electroweak 
symmetry breaking in determining the structure of 
short-distance dominated flavour-changing processes in the SM.

\subsubsection{Hadronic matrix elements}
\label{sect:hqet-scet}

As anticipated, all non-perturbative effects are confined in the hadronic
matrix elements of the operators of the effective Lagrangians. As far as 
the evaluation of the matrix elements is concerned, we can divide 
$B$-physics observables in three main categories: i) inclusive decays, 
ii) one-hadron final states, iii) multi-hadron processes. 

The heavy-quark expansion~\cite{Georgi:1990um}
form a solid theoretical framework to evaluate the 
hadronic matrix elements for inclusive processes: 
inclusive hadronic rates are related 
to those of free $b$ quarks, calculable in perturbation 
theory, by means of a systematic expansion in 
inverse powers of $\Lambda_{\rm QCD}/m_b$. Thanks to 
quark-hadron duality, the lowest-order terms in this 
expansion are the pure partonic rates, and for sufficiently 
inclusive observables higher-order corrections are usually
very small. This technique has been very successful in the past
in the case of charged-current semileptonic decays, as well 
as $B\to X_s\gamma$. However, it has a limited domain of 
applicability, due to the difficulty of selecting and reconstructing
hadronic inclusive states. It cannot be used at hadronic machines,
and even at $B$ factories it cannot be applied to very 
rare decays.

For processes with a single hadron in the final state, the hadronic effects 
are often (although not always) confined to the matrix elements of 
a single quark current. These can be expressed in terms of the meson decay constants
\be 
\langle 0 \vert  b \gamma_\mu \gamma_5 q  \vert  B_q(p) \rangle = 
i p_\mu F_{B_q}~,
\label{eq:FB}
\ee
or appropriate $B\to H$ hadronic form factors. Lattice QCD is the best tool 
to evaluate these non-perturbative quantities from first principles, at least 
in the kinematical region where the form factors are real 
(no re-scattering phase allowed). 
At present not all the form-factors relevant for $B$-physics 
phenomenology are computed on the lattice with good accuracy, 
but the field is evolving rapidly (see e.g.~Ref.~\cite{Gamiz:2008iv,Lubicz:2008am}).
To this category belong also the so-called bag-parameters 
for $\Delta B=2$ mixing, $B_{d,s}$, defined by 
\be 
\eta_B(\mu) \langle \bar B_{q} \vert  (\bar b_L \gamma_\mu q_L)^2 \vert  B_q
\rangle = \frac{2}{3} f_{B_q}^2 m_{B_q}^2 \eta_B(\mu) B_q(\mu) =
\frac{2}{3} f_{B_q}^2 m_{B_q}^2 \hat \eta_B \hat B_q~,
\ee
where both $\hat B_q$ and $\hat \eta_B$ are scale-independent quantities
($\hat \eta_B = 0.55 \pm 0.01$). For later convenience, we report here 
the lattice averages for meson decay constants and bag parameters
obtained in Ref.~\cite{Lubicz:2008am}:
\bea
&& F_{B_s} = 245 \pm 25~{\rm GeV}~, \qquad\quad \hat B_s = 1.22 \pm 0.12~,  \\
&& \frac{F_{B_s}}{F_{B_d}} =1.21 \pm 0.04~, \qquad\qquad \frac{\hat B_{B_s}}{\hat B_{B_d}} = 1.00 \pm 0.03~. 
\eea
As can be seen, the absolute determinations are affected by $\cO(10\%)$ errors, 
while the ratios, which are sensitive to $SU(3)$ breaking effects only, 
are known with a better precision. This is why the ratio
$\Delta m_{B_d}/\Delta m_{B_s}$ gives more significant constraint in Fig.~\ref{fig:1utriangle}
rather than $\Delta m_{B_d}$ only.

The last class of hadronic matrix elements is the one of 
multi-hadron final states, such as the two-body non-leptonic decays $B\to \pi\pi$ and
$B\to K\pi$, as well as many other processes with more than one hadron in the final state. 
These are the most difficult ones to be estimated from first principles with high 
accuracy. A lot of progress in the recent pass has been achieved thanks to 
QCD factorization~\cite{Beneke:2000ry}  and the SCET~\cite{Bauer:2000yr}
approaches, which provide factorization formulae
to relate these hadronic matrix elements to two-body hadronic form factors
in the large $m_b$ limit. 
However, it is fair to say that the errors associated to the $\Lambda_{\rm QCD}/m_b$
corrections are still quite large. 
This subject is quite interesting by itself, but is beyond the scope of these lectures,
where we focus on clean $B$-physics observables for NP studies.
To this purpose, the only interesting non-leptonic channels are those where, 
with suitable ratios, or using $SU(2)$ relations among hadronic matrix elements, 
we can eliminate completely all hadronic unknowns. 
Examples of this type are the $B\to DK$ channels discussed in Sect.~\ref{sect:gamma}.

\subsection{Time evolution and of $B_{d,s}$ states}               
\label{sect:BBmix}


The non vanishing amplitude mixing a $B^0$  meson ($B^0_{d}$ or $B^0_{s}$) 
with the corresponding anti meson, described within the SM by the effective Lagrangian
in Eq.~\ref{eq:db2}, 
induces a time-dependent oscillations between $B^0$ and $\bar B^0$ states:
an initially produced $B^0$ or $\bar B^0$ evolves in time into a
superposition of $B^0$ and $\bar B^0$. Denoting by  $\ket{B^0 (t)} $
(or $\ket{\bar B^0 (t)}$) the
state vector of a $B$ meson which is tagged as a $B^0$ (or $\bar B^0$)
at time $t=0$, the time evolution of these states is governed 
by the following equation:
\be
i \frac{d}{dt} \left(\ba{c} \ket{B(t)} \\ \ket{\bar B(t)} \ea \right)
= \left( M - i\, \frac{\Gamma}{2} \right) 
\left(\ba{c} \ket{B(t)} \\ \ket{ \bar B (t)} \ea \right)~,
\label{mgmat}
\ee
where the mass-matrix $M$ and the decay-matrix $\Gamma$
are $t$-independent, Hermitian $2\times 2$ matrices. 
CPT invariance implies that $M_{11} = M_{22}$ and $\Gamma_{11} = \Gamma_{22}$,
while the off-diagonal element  $M_{12}=M_{21}^*$ is the one we can compute 
using the effective Lagrangian ${\cal L}_{\Delta B=2}$. 

The mass eigenstates are the eigenvectors of $M - i\,\Gamma/2$. We
express them in terms of the flavor eigenstates as
\be
\ket{B_L} = p \ket{B^0} + q \ket{\bar B^0}~,  \qquad 
\ket{B_H} = p \ket{B^0} - q \ket{\bar B^0}~,
\label{defpq}
\ee
with $|p |^2+|q |^2 = 1$.
Note that, in general, $\ket{B_L}$ and $\ket{B_H}$ are not orthogonal 
to each other. The time evolution of the mass eigenstates is 
governed by the two eigenvalues $M_H -i\,\Gamma_H/2$ and $M_L -i\,\Gamma_L/2$:
\be
\ket{B_{H,L} (t) } = e^{-(i M_{H,L} + \Gamma_{H,L}/2)t} \, 
  \ket{B_{H,L}(t=0) } \,. \label{thl}
\ee
For later convenience it is also useful to define 
\be
m = \frac{ M_H + M_L}{2},  \qquad \Gamma =  \frac{\Gamma_L + \Gamma_H}{2}~, \qquad 
\Delta m  =  M_H - M_L~, \qquad  \Delta \Gamma = \Gamma_L - \Gamma_H~.
\label{mg} 
\ee
With these conventions the time evolution of initially tagged $B^0$ or 
$\bar B^0$ states is 
\bea
\ket{B^0 (t)} &=&  e^{-i m t} \, e^{-\Gamma t/2} \left[  
  f_+ (t)\, \ket{B^0} + \frac{q}{p}\, f_- (t)\, \ket{ \bar B^0} \right]~, \no\\
\ket{\bar B^0 (t)} &=& e^{-i m t} \, e^{-\Gamma t/2} \left[ \frac{p}{q}\, f_- (t)\, \ket{B^0} 
  +  f_+(t)\, \ket{\bar B^0} ~\right]~,
  \label{tgg}    
\eea 
where
\bea
f_+ (t) &=& \phantom{-} \cosh\frac{\Delta \Gamma t}{4} \cos\frac{\Delta m t}{2} -   
      i \sinh\frac{\Delta \Gamma t}{4} \sin \frac{\Delta m t}{2}~, \\
f_- (t) &=&  - \sinh\frac{\Delta \Gamma t}{4} \cos \frac{\Delta m t}{2} +
      i \cosh\frac{\Delta \Gamma t}{4} \sin \frac{\Delta m t}{2}~,  
\label{gpgm} 
\eea

In both $B_s$ and $B_d$ systems the following hierarchies holds: 
$|\Gamma_{12}| \ll | M_{12}|$ and $\Delta \Gamma \ll \Delta m$.
They are experimentally verified and can be traced back 
to the fact that $|\Gamma_{12}|$ is a genuine long-distance $\cO(G_F^2)$ effect
(it is indeed related to the absorptive part of the box diagrams 
in Fig.~\ref{fig:BBmix}) which do not share the large $m_t$ enhancement of 
 $|M_{12}|$ (which is a short-distance dominated quantity).  
Taking into account this hierarchy leads to the following 
approximate expressions for the quantities appearing in the time-evolution formulae 
in terms of $M_{12}$ and $\Gamma_{12}$:
\bea
\Delta m &=& 2\, |M_{12}| \left[ 1 + 
  {\cal O} \left( \left| \frac{\Gamma_{12}}{M_{12}} \right|^2 \right) \right]~,
  \label{mgsol:a} \\
\Delta \Gamma &=& 2\, |\Gamma_{12}| \cos \phi \left[ 1+  
  {\cal O} \left( \left| \frac{\Gamma_{12}}{M_{12}} \right|^2 \right) \right]~,
  \label{mgsol:b} \\
\frac{q}{p} &=& - e^{- i \phi_B} \left[ 1 - \frac{1}{2}\left| \frac{\Gamma_{12}}{M_{12}} \right|\sin\phi
   + {\cal O} \left( \left| \frac{\Gamma_{12}}{M_{12}} \right|^2 \right) \right]~,
\label{qpsol}  
\eea
where $\phi = {\rm arg}( - M_{12}/\Gamma_{12})$ and $\phi_B$ is the phase of $M_{12}$.
Note that $\phi_B$ thus defined is not measurable and 
depends on the phase convention adopted, while $\phi$ is a phase-convention quantity which can be measured in
experiments.

Taking into account the above results, the time-dependent decay rates 
of an initially tagged $B^0$ or $\bar B^0$ state
into some final state $f$ can be written as
\bea
&& \Gamma[B^0(t=0) \to f(t)]  =  {\cal N}_0  | A_f |^2  e^{-\Gamma t}
  \Bigg\{ \frac{1 + | \lambda_f |^2}2 \cosh \frac{\Delta \Gamma  t}{2} \no \\
&& \qquad 
+   \frac{ 1 - | \lambda_f |^2}2 \cos ( \Delta m  t )  
  - \Re \lambda_f  \sinh \frac{\Delta \Gamma  t}{2} 
  - \Im \lambda_f  \sin ( \Delta m  t) \Bigg\} , \no \\
&& \!\!\!\!\! \Gamma[\bar B^0(t=0) \to f(t)] = {\cal N}_0  | A_f |^2  
\left( 1 + \left| \frac{\Gamma_{12}}{M_{12}} \right|\sin\phi  \right) 
  e^{-\Gamma t} \Bigg\{ \frac{1 + | \lambda_f |^2}2 \times \no \\
 && \times  \cosh \frac{\Delta \Gamma  t}{2}  
   - \frac{1 - | \lambda_f |^2}2 \cos ( \Delta m  t )   
    - \Re \lambda_f  \sinh \frac{\Delta \Gamma  t}{2} 
    + \Im \lambda_f  \sin ( \Delta m  t ) \Bigg\}~, \no
\eea
where ${\cal N}_0$ is the flux normalization and, following the standard notation, 
we have defined 
\be
\lambda_f = \frac{q}{p} \frac{\bar A_f}{A_f}
  \approx - e^{- i \phi_B}\, \frac{\bar A_f}{A_f}  \left[ 1 
- \frac{1}{2}\left| \frac{\Gamma_{12}}{M_{12}} \right|\sin\phi \right] 
\ee
in terms of the decay amplitudes 
\be 
A_f = \langle f | \cL_{\Delta F=1} \ket{B^0}~, \qquad \qquad \bar A_f = \langle f |
 \cL_{\Delta F=1} \ket{\bar B^0}~. \label{defaf}
\ee
From the above expressions it is clear that the key 
quantity we can access experimentally in the time-dependent study of $B$ decays 
is the combination $\lambda_f$. Both real and imaginary parts of  $\lambda_f$
can be measured, and indeed this is a phase-convention  independent quantity: the phase 
convention in $\phi_B$ is compensated by the phase convention in the decay
amplitudes. In other words, what we can measure is the weak-phase difference 
between $M_{12}$ and the decay amplitudes.

For generic final states, $\lambda_f$ is a quantity that is difficult to evaluate.
However, it becomes particularly simple in the case where $f$ is a CP eigenstate,
${\rm CP}\ket{f} = \eta_f \ket{f}$, and the weak phase of the decaying 
amplitude is know. In such case  $\bar A_f/A_f$ is a pure phase factor ($|\bar A_f / A_f|=1$), 
determined by the weak phase of the decaying amplitude:
\be
\left. \lambda_f \right|_{\rm CP-eigen.} 
= \eta_f \frac{q}{p} e^{-2i \phi_{A} }~, \qquad A_f =| A_f| e^{i \phi_A}~, \qquad \eta_f=\pm 1~.
\ee
The most clean example of this type of channels is the $\ket{\psi K_S}$ final state for 
$B_d$ decays. In this case the final state is a CP eigenstate and the decay amplitude 
is real (to a very good approximation) in the standard CKM phase convention. 
Indeed the underlying partonic transition is dominated by the Cabibbo-allowed 
tree-level process $b\to c \bar c s$, which has a vanishing phase in the 
standard CKM phase convention, and also the leading one-loop corrections 
(top-quark penguins) have the same vanishing weak phase.
Since in the $B_d$ system we can safely neglect $\Gamma_{12}/M_{12}$,
this implies
\be
\lambda^{B_b}_{\psi K_s} = - e^{- i \phi_{B_d} }~, 
\qquad \Im\left(\lambda^{B_b}_{\psi K_s}\right)_{\rm SM} = \sin(2\beta) ~,
\ee
where the SM expression of $\phi_{B_d}$ 
is nothing but the phase of the CKM combination $(V_{tb}^* V_{td})^2$ appearing in 
Eq.~(\ref{eq:db2}). Given the smallness of $\Delta \Gamma_d$, 
this quantity is easily extracted by the ratio
\bea
\frac{ \Gamma[\bar B_d(t=0) \to \psi K_s (t)] - \Gamma[B^0(t=0) \to f \psi K_s (t) ] }
{ \Gamma[\bar B_d(t=0) \to \psi K_s (t)] + \Gamma[B^0(t=0) \to f \psi K_s (t) ] }
 \approx  \Im\left(\lambda^{B_b}_{\psi K_s}\right) \sin(\Delta m_{B_d} t)~,
\no
\eea
which can be considered  the golden measurement of $B$ factories.

Another class of interesting final states are CP-conjugate channels $\ket{f}$
and $\ket{\bar f}$ which are accessible only to $B^0$ or $\bar B^0$ states, 
such that $|A_f| = |\bar A_{\bar f}|$ and  $\bar A_f = \bar A_{f}$=0. Typical examples 
of this type are the charged semileptonic channels. In this case the asymmetry 
\bea
\frac{ \Gamma[\bar B^0(t=0) \to  f (t)] - \Gamma[B^0(t=0) \to \bar f (t) ] }{ 
\Gamma[\bar B^0(t=0) \to f (t)] + \Gamma[B^0(t=0) \to \bar f  (t) ] }
= \left|\frac{\Gamma_{12}}{M_{12}}\right| \sin\phi \left[ 1  
+ {\cal O} \left( \frac{\Gamma_{12}}{M_{12}} \right) \right] 
\no
\eea
turns out to be time-independent and a clean way to determine the indirect
CP-violating phase $\phi$.


\subsection{A selection of particularly interesting observables in the LHC era}

Given the general arguments about the sensitivity to physics beyond the SM presented 
at the end of the first lecture, and the theoretical tool discussed above, in the following
we analyse some interesting measurements which could be performed in the LHC
era, and particularly at Tevatron and at the LHCb experiment. 
The list presented here is far from being exhaustive (for a more complete analysis
we refer to the reviews in Ref.~\cite{Antonelli:2009ws,Buchalla:2008jp,Buras:2009if,Anikeev:2001rk}),
but it should serve as an illustration of the interesting potential of $B$ 
physics at hadron colliders.

\subsubsection{CP violation  in $B_s$ mixing}

The CP violating phase of  $B_s$--$\bar B_s$  mixing is the last missing 
ingredient of down-type $\Delta F=2$ observables. The fact we have not 
found any deviations from the SM in $B_d$--$\bar B_d$  and $\Delta m_{B_s}$
should not discourage the measurement of this missing piece of the
$\Delta F=2$ puzzle. 

\begin{figure}[t]
  \centering
  {\includegraphics[width=0.5\textwidth]{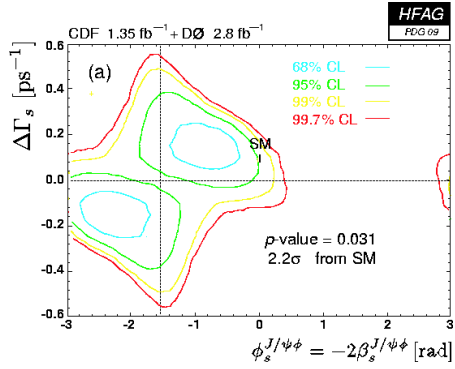}}
 \caption{Combined fit of CDF and D0 results on $\Delta \Gamma$ and $\phi_{B_s}$ from the 
time-dependent analysis of  $B_s \to \psi \phi$ decays, from Ref.~\cite{Barberio:2008fa}. }
  \label{fig:phis}
\end{figure}

The golden channel for the measurement of  the CP-violating phase of  
$B_s$--$\bar B_s$ mixing is the time-dependent analysis  
of the $B_s(\bar B_s) \to \psi \phi$ decay. At the quark level 
$B_s \to \psi \phi$  share the same virtues of  $B_d \to \psi K$ 
(partonic amplitude of the type  $b\to c \bar c s$), 
which is used to extract the phase of $B_d$--$\bar B_d$ mixing.
However, there a few points which makes this measurement much 
more challenging:
\begin{itemize}
\item The $B_s$ oscillations are much faster ($\Delta m_{B_s}/\Delta m_{B_d} 
\approx F^2_{B_s}/F^2_{B_d} | V_{ts}/V_{td}|^2$ $\approx 30$), making the time-dependent analysis
quite difficult (and essentially inaccessible at $B$ factories).
\item Contrary to $\ket{\psi K}$, which has a single angular momentum and is a pure CP eigenstate, 
the vector-vector state $\ket{\psi \phi}$ produced by the $B_s$ decay has different angular momenta, 
corresponding to different CP eigenstates. These must be disentangled with a proper angular 
analysis of the final four-body final state $\ket{(\ell^+\ell^-)_{\psi} (K^+K^-)_{\phi}}$.
To avoid contamination from the nearby $\ket{\psi f_0}$ state, the fit should include also
a $\ket{(\ell^+\ell^-)_{\psi} (K^+K^-)_{S-{\rm wave}}}$ component, for a total of ten independent
(and unknown) weak amplitudes.\footnote{~The formalism is essentially the same 
adopted in the four-body angular analysis of the $B_d \to J/\psi K \pi$ decay at
$B$ factories~\cite{Aubert:2004cp}, with the only difference that $\Delta \Gamma$
cannot be neglected in the $B_s$ case.}
\item Contrary to the $B_d$ system, the width difference cannot be neglected in the 
$B_s$ case, leading to an additional key parameter to be included in the fit. 
\end{itemize}
Modulo the experimental difficulties listed above, the process is theoretically clean and 
a complete fit of the decay distributions should allow the extraction of 
\be
\lambda^{B_s}_{\psi \phi} \approx - e^{- i \phi_{B_s} }~, 
\ee
where the SM prediction is 
\be 
\phi_{B_s}^{\rm SM}  = - {\rm arg} \frac{ (V_{tb}^* V_{ts})^2 }{ |V_{tb}^* V_{ts}|^2 } \approx  -0.03~.
\ee
The tiny value of $\phi^{\rm SM}_{B_s}$ implies that, within the SM, 
no CP asymmetry should be observed in the near future.  The present status of the combined 
fit of $\Delta \Gamma_s$ and $\phi_{B_s}$ from the CDF~\cite{Aaltonen:2007he} and D0~\cite{:2008fj}
 experiments at Tevatron is shown in Fig.~\ref{fig:phis}. 
As can be noted, the agreement with the SM is not good, but the 
errors are still large.

As we will see in Sect.~\ref{sect:MFV} a clear evidence for $\phi_{B_s}\not =0$
at Tevatron, or even within the first year of LHCb
(which realistically would imply $|\phi_{B_s} | > 0.1$), 
would not only signal the presence of physics beyond the SM, but would 
also rule out the whole class of MFV models.

\subsubsection{CP violation in charged $B$ decays}
\label{sect:gamma}

Among non-leptonic channels $B^\pm \to D K$ decays
are particularly interesting since, via appropriate asymmetries, 
allows us to extract the CKM angle $\gamma$ in a very clean way.
The extraction of $\gamma$ involves only tree-level $B$ decay 
amplitudes, and is virtually free from hadronic uncertainties
(which are eliminated directly by data). It is therefore an
essential element for a precise determination of the SM Yukawa
couplings also in presence of NP.

The main strategy is based on the following two observations:
\begin{itemize}
\item{} The partonic amplitudes for  $B^- \to \bar D K^-$ ($b \to c \bar u s$) 
and $B^- \to \bar D K^-$ ($b \to u \bar c s$) are pure tree-level 
amplitudes (no penguins allowed given the four different quark flavours).
As a result, their weak phase difference is completely determined and
is $\gamma = {\rm arg}\left(-V_{ud} V^*_{ub}/V_{cd}V^*_{cb}\right)$.
\item{} Thanks to $D$--$\bar D$  mixing, there are several 
final states $f$ accessible to both $D$ and $\bar D$, where the 
two tree-level amplitudes can interfere. By combining 
the four final states $B^\pm \to f K^\pm$ and $B^\pm \to \bar f K^\pm$,
we can extract $\gamma$ and all the relevant hadronic unknowns
of the system. 
\end{itemize}
The first strategy, proposed by Gronau, London, and Wyler~\cite{Gronau:1990ra}
was based on the selection of $D(\bar D)$ decays to two-body 
$CP$ eigenstates. But it has later been realized that any final state
accessible to both  $D$ and $\bar D$ (such as the $K^\pm\pi^\mp$ channels~\cite{Atwood:1996ci}, 
or multibody final states~\cite{Giri}) may work as well.

Let's start from the case of $D(\bar D)$ decays to CP eigenstates,
where the formalism is particularly transparent. The key quantity is 
the ratio 
\begin{equation}
  r_B e^{i \delta_B} = \frac{ A(B^+ \to D^0 K^+)  }{ A(B^+ \to \bar D^0 K^+)}~,
\end{equation}
where $\delta$ is a strong phase. Denoting CP-even and CP-odd final 
states $f_+$ and $f_-$, we then have  
\bea
  {A}\left(B^{-}\to f_{+}K^{-}\right) &=& 
A_0 \times \left[1+r_{B}e^{i \left(\delta_B - \gamma\right)}\right] \no \\
  {A}\left(B^{-}\to f_{-}K^{-}\right) &=& 
A_0  \times\left[1-r_{B}e^{i \left(\delta_B - \gamma\right)}\right] \no \\
  {A}\left(B^{+}\to f_{+}K^{+}\right)  &=& 
A_0 \times \left[1+r_{B}e^{i \left(\delta_B + \gamma\right)}\right] \no \\
  {A}\left(B^{+}\to f_{-}K^{+}\right)  &=& 
A_0  \times\left[1-r_{B}e^{i \left(\delta_B + \gamma\right)}\right] 
\eea
It is clear that combining the four rates we can extract
the three hadronic unknowns ($A_0$, $r_{B}$, and $\delta$) as well as $\gamma$.
It is also clear that the sensitivity to $\gamma$ vanishes 
in the limit $r_B \to 0$, and indeed the main limitation of this method 
is that $r_B$ turns out to be very small.

The formalism is essentially unchanged if we consider final states 
that are not CP eigenstates, such as the $K^\pm\pi^\mp$ states. These
have the advantage that the suppression of  $r_B$
is partially compensated by the CKM suppression of the 
corresponding $D(\bar D)\to K^\pm\pi^\mp$ decays. Indeed 
the effective relevant ratio becomes
\begin{equation}
  r_{\rm eff} e^{i \delta_{\rm eff}} =  \frac{ A(B^+ \to D^0 K^+)  }{ A(B^+ \to \bar D^0 K^+)}~,
\times 
\frac{A(D^0\to K^-\pi^+)}{A(\bar D^0\to K^-\pi^+)}
\end{equation}
which is substantially larger than $r_B$.

Once  $r_B$ and $\delta_B$ (or $r_{\rm eff}$ and $\delta_{\rm eff}$) are determined from data,
the extraction of $\gamma$  has essentially no theoretical
uncertainty. In principle a theoretical error could be induced by the neglected 
CP-violating effects in charm mixing. In practice, the experimental bounds on 
charm mixing make this effect totally negligible. 
The key issue is only collecting high statistics on this highly-suppressed decay 
modes: a clear target for $B$ physics at hadron machines. 

\subsubsection{The forward-backward asymmetry in $B\to K^* \ell^+\ell^-$.}
\label{sect:BKll} 

Theoretical predictions for exclusive FCNC decays are not easy.
Even if the final state involve only one hadron, in most of
the kinematical region re-scattering effects of the type
$B\to K^* H \bar H \to K^*  \ell^+\ell^-$ are possible,
making difficult to estimate precisely the decay rate.
However, the largest source of uncertainty is typically 
the normalization of the hadronic form factors. The theoretical 
error can be substantially reduced in appropriate 
ratios or differential distributions. A clean example of this
type is the normalized forward-backward asymmetry in $B\to K^* \ell^+\ell^-$. 

\begin{figure}[t]
\begin{center}
  \includegraphics[height=.21\textheight]{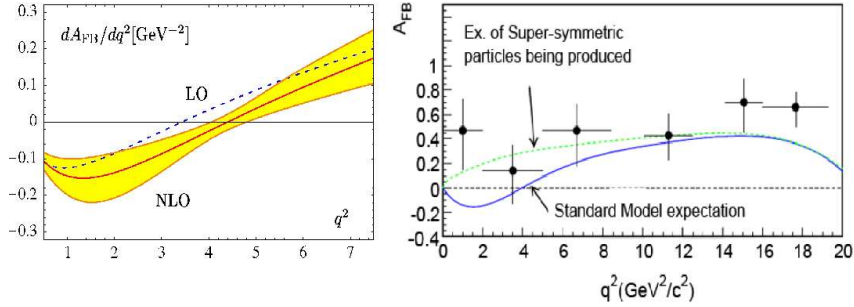}
\end{center}
\vskip -0.5 cm
\caption{Left: SM prediction for the zero of the forward-backward asymmetry 
in $B^- \to K^{*-} \ell^+\ell^-$ at LO and NLO (the band indicates the  
theoretical uncertainty)~\cite{Martin}. Right: Recent measurement of  the forward-backward asymmetry 
in $B^- \to K^{*-} \ell^+\ell^-$ by BELLE~\cite{:2009zv}} 
\label{fig:martin}
\end{figure}

The observable is defined as 
\be
\label{eq:asdef}
\cA_{FB}(s)=\frac{1}{d\Gamma(B\to  K^* \mu^+\mu^- )/ds}
  \int_{-1}^1 d\cos\theta ~
 \frac{d^2 \Gamma(B\to  K^* \mu^+\mu^- )}{d s~ d\cos\theta}
\mbox{sgn}(\cos\theta)~,
\ee
where $\theta$ is the angle between the momenta of 
$\mu^+$ and $\bar B$ in the dilepton center-of-mass frame. 
Assuming that the leptonic current has only a 
vector ($V$) or axial-vector ($A$) structure
(as in the SM), the FB asymmetry provides a direct measure of 
the $A$--$V$ interference. Indeed, at the lowest-order one can write
$$
{\cal A}_{\rm FB}(q^2)
 \propto 
  {\rm Re}\left\{  C_{10A}^* \left[ \frac{q^2}{m_b^2} C_{9V}^{\rm eff} 
    + r(q^2) \frac{m_b C_{7\gamma}}{m_B}  \right] \right\}~,
$$
where $r(q^2)$ is an appropriate ratio of $B\to K^*$
vector and tensor form factors~\cite{burdman0}. 
There are three main features of this observable
that provide a clear and independent 
short-distance information: 
\begin{enumerate}
\item[1.]~The position of the zero ($q_0$) of ${\cal A}_{\rm FB}(q^2)$ in the 
low-$q^2$ region (see Fig.~\ref{fig:martin}) \cite{burdman0}:
as shown by means of a full NLO calculation \cite{Martin}, 
the experimental measurement of $q^2_0$ could 
allow a determination of $C_7/C_9$ at the  $10\%$ level.
\item[2.]~The sign of ${\cal A}_{\rm FB}(q^2)$ around the zero.
This is fixed unambiguously in terms of the relative sign
of $C_{10}$ and $C_9$: within the SM one 
expects ${\cal A}_{\rm FB}(q^2 > q^2_0) > 0$  for 
$|\bar B \rangle \equiv |b \bar d \rangle$ mesons.
\item[3.]~The relation ${\cal A}[\bar B]_{\rm FB}(q^2) = - {\cal A}[B]_{\rm FB}(q^2)$.
This follows from the CP-odd structure of ${\cal A}_{\rm FB}$
and holds at the $10^{-3}$ level within the SM \cite{BHI}, 
where $C_{10}$ has a negligible CP-violating phase.
\end{enumerate}

The present status of the measurement at BELLE~\cite{:2009zv} is shown in Fig.~\ref{fig:martin}.
Also in this case, similarly to $B_s$--$\bar B_s$ mixing, the agreement with the SM is not good,
leaving open the room for speculations about sizable non-standard effects.
However, the errors are clearly too large to draw definite conclusions.

\subsubsection{$B\to \ell^+\ell^-$}
\label{sect:Bll} 

The purely leptonic decays constitute a 
special case among exclusive transitions. Within the SM 
only the axial-current operator, $Q_{10A}$, 
induces a non-vanishing contribution 
to these decays. As a result, the short-distance contribution
is not {\em diluted} by the mixing with four-quark operators.
Moreover, the hadronic matrix element involved is the simplest
we can consider, namely the $B$-meson decay constant in Eq.~(\ref{eq:FB}).
As we have seen, 
present estimates of $F_{B_d}$  and $F_{B_s}$ from lattice QCD are already
at the $10\%$ level, and in the future the error is expected 
to go below $5\%$.

The price to pay for this theoretically-clean amplitude is a strong helicity 
suppression for $\ell=\mu$ (and $\ell=e$), or the channels 
with the best experimental signature. Employing the 
full NLO expression  of $C^{\rm SM}_{10A}$,
we can write~\cite{Buras:2003td}
\bea
&&{\mathcal B}( B_s \to \mu^+ \mu^-)^{\rm SM} = 3.1 \times 10^{-9} 
\left( \frac{|V_{ts}|}{0.04}  \right)^2 \times \left( \frac{F_{B_s}}{0.21~\mbox{GeV}} \right)^2 \!\!
\label{eq:BrmmSM} \\ 
&&\frac{ {\mathcal B}( B_s \to \tau^+ \tau^-)^{\rm SM} }{  
{\mathcal B}( B_s \to \mu^+ \mu^-)^{\rm SM} } = 215~, \qquad 
\frac{ {\mathcal B}( B_s \to e^+ e^-)^{\rm SM} }{  
{\mathcal B}( B_s \to \mu^+ \mu^-)^{\rm SM} } = 2.4 \times 10^{-5}~.
\eea
The corresponding $B_d$ modes are both 
suppressed by an additional factor $|V_{td}/V_{ts}|^2  F^2_{B_d}/F^2_{B_s}
 \approx 1/30$. 
The present experimental bound closest to SM expectations is the one obtained by 
CDF on $B_{s}\to \mu^+ \mu^-$~\cite{Aoki:2009gv}
\be
{\mathcal B}(B_{s}\to \mu^+ \mu^-) < 5.8 \times 10^{-8} \quad (95 \% \;\;  {\rm CL})~,
\ee
which is about one order of magnitude above the SM level.

The strong helicity suppression and the theoretical cleanness
make these modes excellent probes of several new-physics
models and, particularly, of scalar FCNC amplitudes. 
Scalar FCNC operators, such as $\bar b_R s_L \bar \mu_R \mu_L$, 
are present within the SM but are  
negligible because of the smallness 
of down-type Yukawa couplings. On the other hand,
these amplitudes could be non-negligible
in models with an extended Higgs sector (see Sect.~\ref{eq:largetanb}).
In particular, within the MSSM, where two Higgs doublets are 
coupled separately to up- and down-type quarks, a strong 
enhancement of scalar FCNCs can occur 
at large $\tan\beta = v_u/v_d$.
This effect is very small in non-helicity-suppressed 
$B$ decays (because of the small 
Yukawa couplings), but could easily enhance $B\to \ell^+\ell^-$
rates by one order of magnitude.

\begin{figure}[t]
\begin{center}
  \resizebox{5.7cm}{!}{\includegraphics{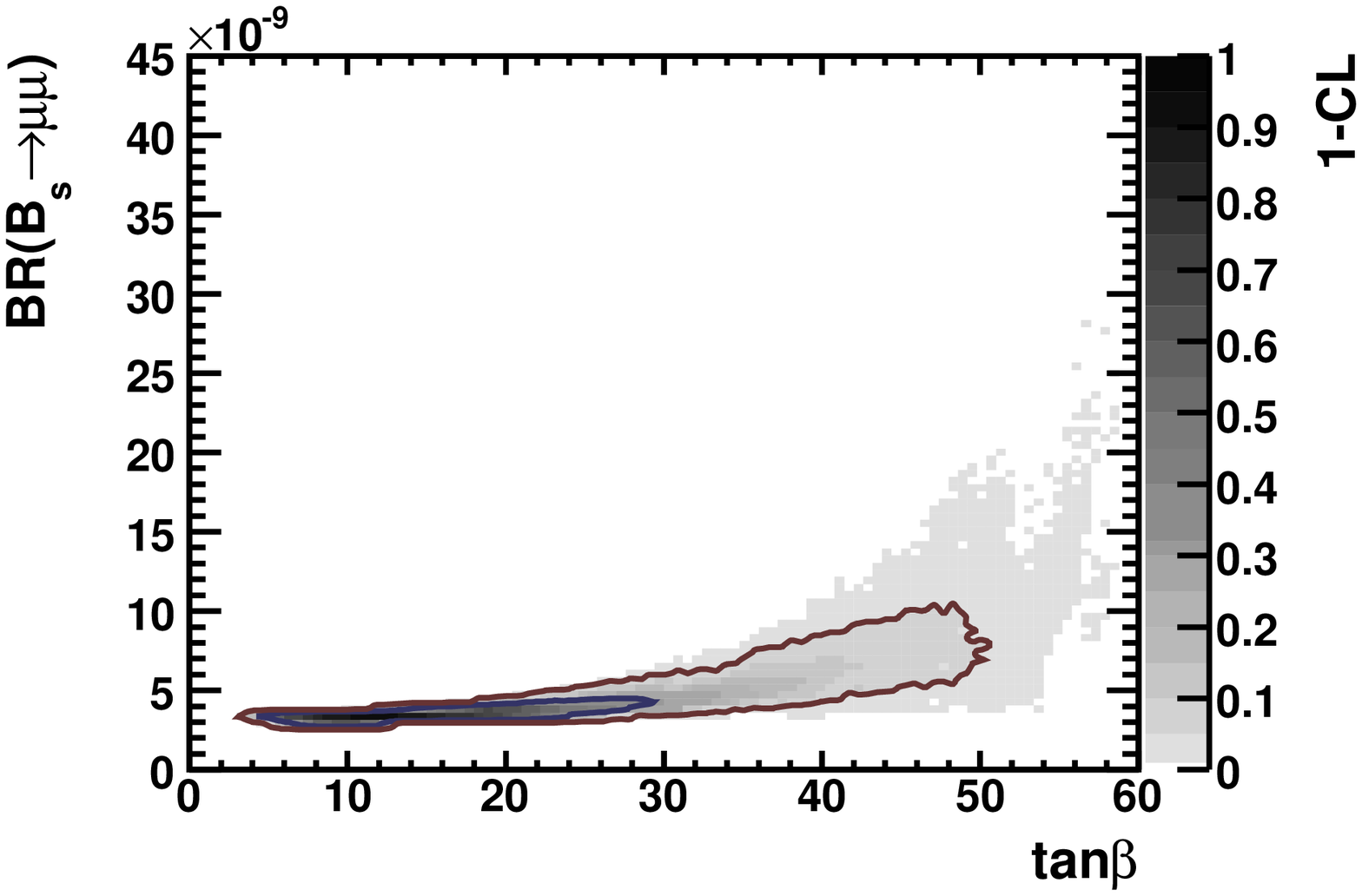}}
  \resizebox{5.7cm}{!}{\includegraphics{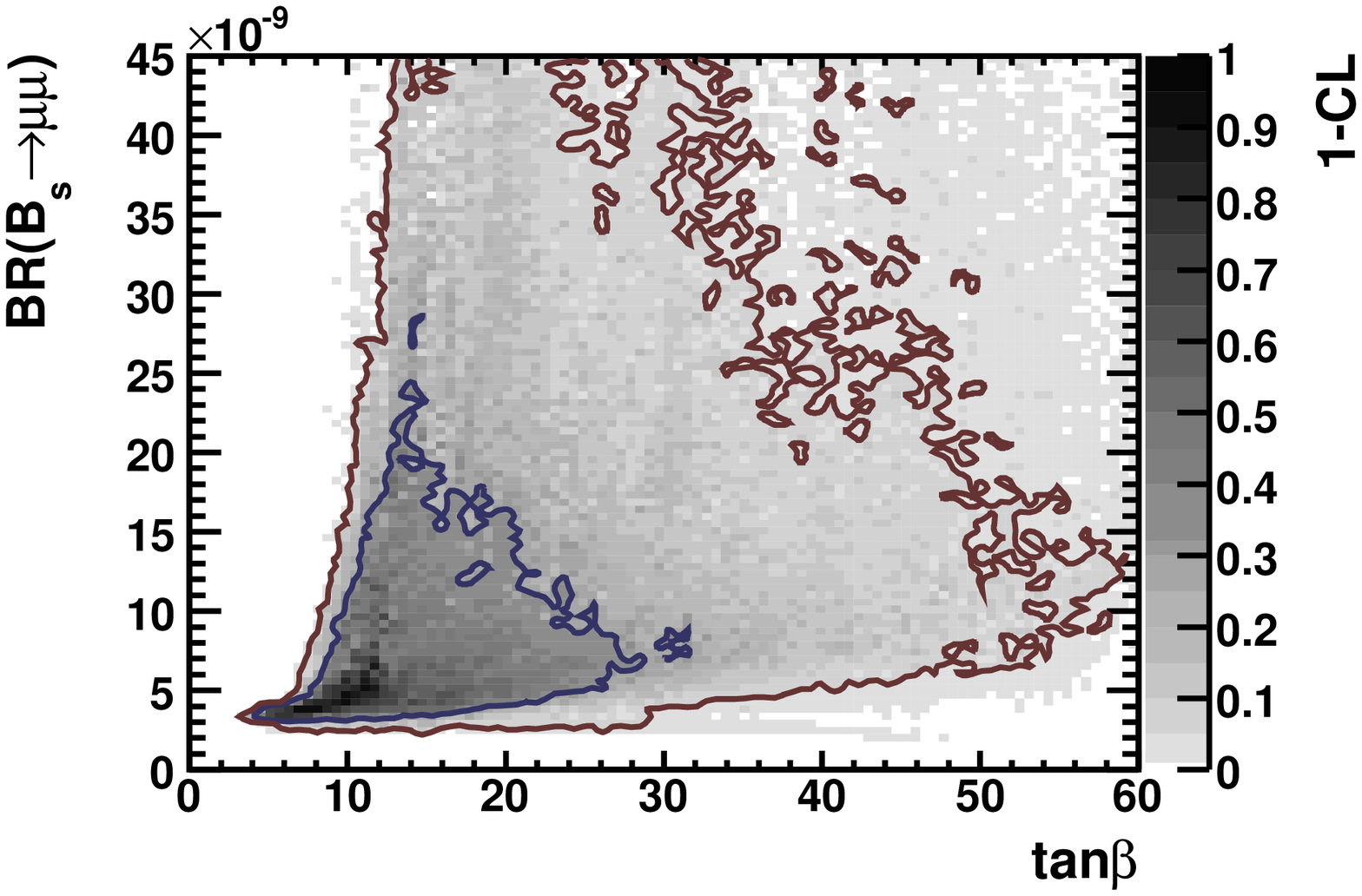}}
\end{center}
\vskip -0.5 cm 
\caption{\it The correlation between branching ratio for $B_{s}\to \mu^+ \mu^-$
and $\tan \beta$ in the CMSSM (left panel) and in the CMSSM
with non-universal Higgs masses (right panel, see Sect.~\ref{sect:CMSSM}
for more details), from Ref.~\cite{Buchmueller:2009fn}. }
\label{fig:tbbsmumu}
\end{figure}

An illustration of the possible enhancement in a 
constrained version of the MSSM (that will be discussed in 
Sect.~\ref{sect:CMSSM}) is shown in Fig.~\ref{fig:tbbsmumu}.
This figure shows that the present search for $B_s \to \mu^+\mu^-$ 
at CDF is already quite interesting, even if the sensitivity 
is well above the SM level. In a long-term perspective, 
the discovery and precise measurement of all the 
accessible $B \to \ell^+\ell^-$ 
channels is definitely one of 
the most interesting items of the $B$-physics program 
at hadron colliders.

\section{Flavour physics beyond the SM: models and predictions}

\subsection{Minimal Flavour Violation}
\label{sect:MFV}

The main idea of MFV is that flavour-violating 
interactions are linked to the
known structure of Yukawa couplings also beyond the SM. 
In a more quantitative way, the MFV construction consists 
in identifying the flavour symmetry and symmetry-breaking structure 
of the SM and enforce it also beyond the SM.

The MFV hypothesis consists of two ingredients~\cite{D'Ambrosio:2002ex}: 
(1)~a {\em flavor symmetry} and (ii)~a set of {\em symmetry-breaking 
terms}.  The symmetry is noting but the large global 
symmetry ${\mathcal G}_{\rm flavour}$
of the SM Lagrangian in absence of Yukawa couplings
shown in Eq.~(\ref{eq:Gtot}). Since this global symmetry, and particularly 
the ${SU}(3)$ subgroups controlling quark flavor-changing 
transitions, is already broken within the SM, we cannot promote 
it to be an exact symmetry of the NP model. Some breaking would
appear  at the quantum level because of the SM Yukawa interactions.
The most restrictive assumption we can make to {\em protect} in a consistent 
way quark-flavor mixing 
beyond the SM is to assume that $Y_d$ and $Y_u$ are the only 
sources of flavour symmetry  breaking also in the NP model.
To implement and interpret this hypothesis in a consistent way, 
we can assume that ${\mathcal G}_{q}$ is a good symmetry and 
promote $Y_{u,d}$ to be non-dynamical fields ({\em spurions}) with 
non-trivial transformation properties under  ${\mathcal G}_{q}$:
\begin{equation}
Y_u \sim (3, \bar 3,1)~,\qquad
Y_d \sim (3, 1, \bar 3)~.\qquad
\end{equation}
If the breaking of the symmetry occurs at very high energy scales, 
at low-energies we would only be sensitive to the background values of 
the $Y$, i.e.~to the ordinary SM Yukawa couplings. 
The role of the Yukawa in breaking the flavour symmetry becomes 
similar to the role of the Higgs in the the breaking of the 
gauge symmetry. However, in the case of the Yukawa we don't 
know (and we do not attempt to construct) a dynamical 
model which give rise to this symmetry breaking.

Within the effective-theory approach 
to physics beyond the SM introduced in Sect.~\ref{sect:effth}, 
we can say that an effective theory satisfies the criterion of
Minimal Flavor Violation in the quark sector 
if all higher-dimensional operators,
constructed from SM and $Y$ fields, are invariant under CP and (formally)
under the flavor group ${\mathcal G}_{q}$~\cite{D'Ambrosio:2002ex}.

According to this criterion one should in principle 
consider operators with arbitrary powers of the (dimensionless) 
Yukawa fields. However, a strong simplification arises by the 
observation that all the eigenvalues of the Yukawa matrices 
are small, but for the top one, and that the off-diagonal 
elements of the CKM matrix are very suppressed. 
Working in the basis in Eq.~(\ref{eq:Ydbasis}) we have 
\be
\left[  Y_u (Y_u)^\dagger \right]^n_{i\not = j} ~\approx~ 
y_t^n V^*_{it} V_{tj}~.
\ee
As a consequence, in the limit where we neglect light quark masses,
the leading $\Delta F=2$ and $\Delta F=1$ FCNC amplitudes get exactly 
the same CKM suppression as in the SM: 
\begin{eqnarray}
  \mathcal{A}(d^i \to d^j)_{\rm MFV} &=&   (V^*_{ti} V_{tj})^{\phantom{a}} 
 \mathcal{A}^{(\Delta F=1)}_{\rm SM}
\left[ 1 + a_1 \frac{ 16 \pi^2 M^2_W }{ \Lambda^2 } \right]~,
\\
  \mathcal{A}(M_{ij}-\bar M_{ij})_{\rm MFV}  &=&  (V^*_{ti} V_{tj})^2  
 \mathcal{A}^{(\Delta F=2)}_{\rm SM}
\left[ 1 + a_2 \frac{ 16 \pi^2 M^2_W }{ \Lambda^2 } \right]~.
\label{eq:FC}
\end{eqnarray}
where the $\mathcal{A}^{(i)}_{\rm SM}$ are the SM loop amplitudes 
and the $a_i$ are $\mathcal{O}(1)$ real parameters. The  $a_i$
depend on the specific operator considered but are flavor 
independent. This implies the same relative correction 
in $s\to d$, $b\to d$, and  $b\to s$ transitions 
of the same type: a key prediction which can be tested 
in experiment.

\begin{figure}[t]
\begin{center}
\includegraphics[width=57mm]{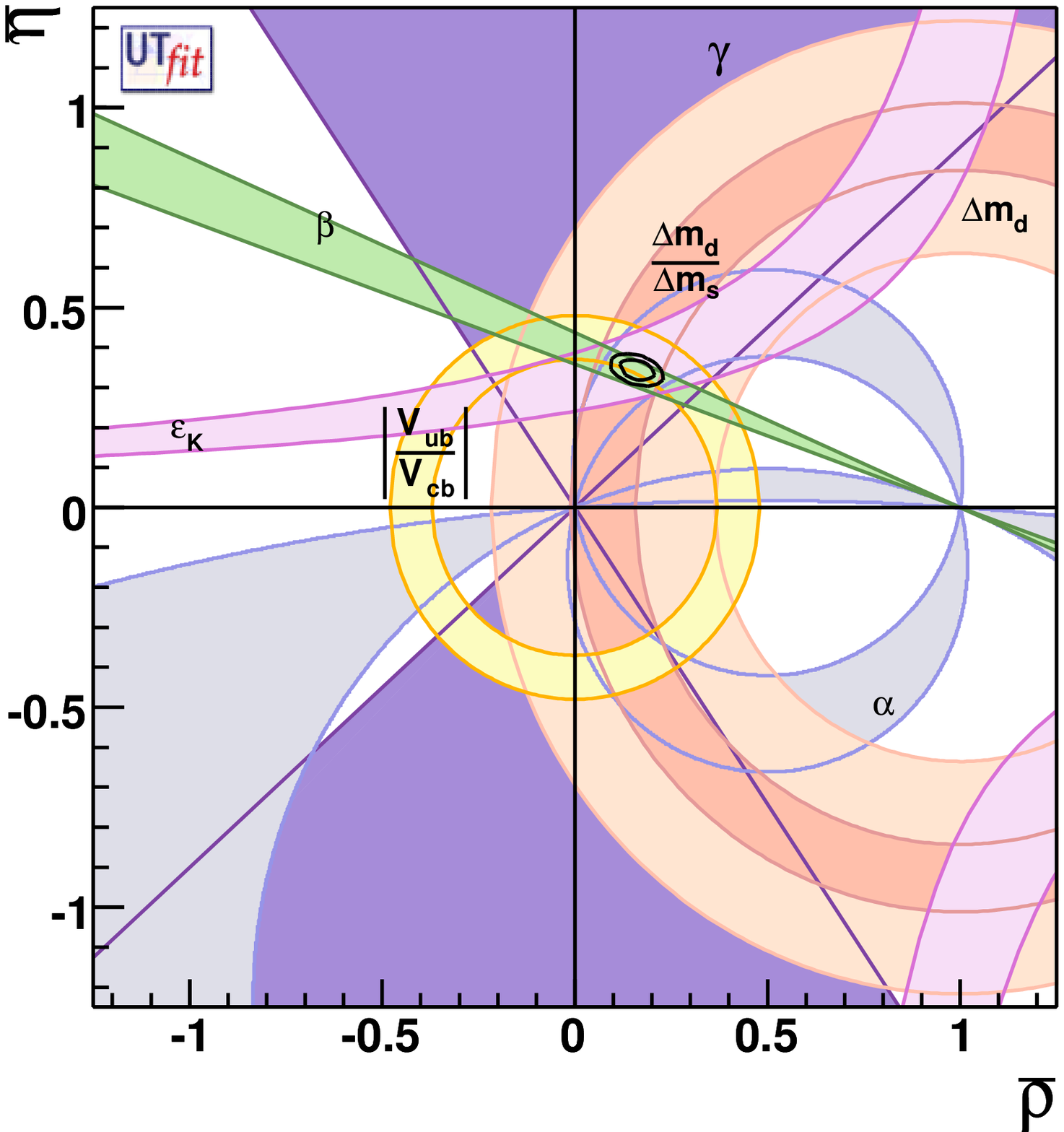}
\includegraphics[width=57mm]{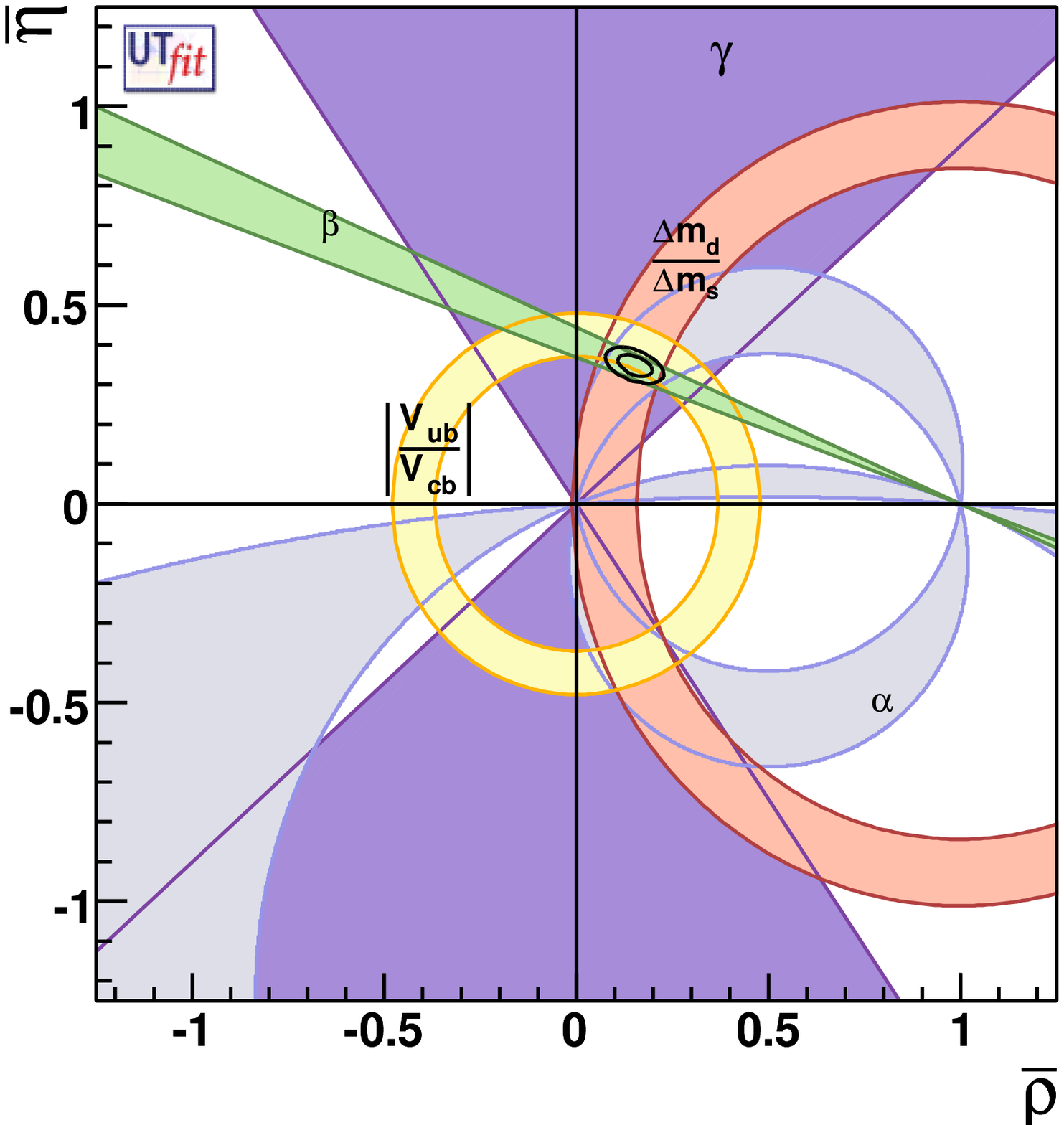}
\caption{\label{fig:UTfits} Fit of the CKM unitarity 
triangle (in 2008) within the SM (left) and 
in generic extensions of the SM 
satisfying the MFV hypothesis (right)~\cite{Bona:2007vi}. }
\end{center}
\end{figure}

As pointed out in Ref.~\cite{Buras:2000dm}, within the MFV
framework several of the constraints used to determine the CKM matrix
(and in particular the unitarity triangle) are not affected by NP.
In this framework, NP effects are negligible not only in tree-level
processes but also in a few clean observables sensitive to loop
effects, such as the time-dependent CPV asymmetry in $B_d \to \psi
K_{L,S}$. Indeed the structure of the basic flavor-changing coupling
in Eq.~(\ref{eq:FC}) implies that the weak CPV phase of $B_d$--$\bar
B_d$ mixing is arg[$(V_{td}V_{tb}^*)^2$], exactly as in the SM.  
This construction provides a natural (a posteriori) justification 
of why no NP effects have been observed in
the quark sector: by construction, most of the clean observables
measured at $B$ factories are insensitive to NP effects in the MFV
framework. A comparison of the CKM fits in the SM and 
in generic MFV models is shown in Fig.~\ref{fig:UTfits}.
Essentially only $\epsilon_K$ and $\Delta m_{B_d}$ (but not 
the ratio  $\Delta m_{B_d}/\Delta m_{B_s}$) are sensitive to 
non-standard effects within MFV models.

Given the built-in CKM suppression, the bounds on 
higher-dimen\-sio\-nal operators in the MFV framework
turns out to be in the TeV range. 
This can easily be understood by the discussion 
in Sect.~\ref{sect:DF2bounds}: the MFV bounds 
on operators contributing to $\epsilon_K$ and $\Delta m_{B_d}$ 
are obtained from  
Eq.~(\ref{eq:boundsDF2}) setting $|c_{ij}| =  |y_t^2 V_{3i}^* V_{3j}|^2$.
From this perspective we could say that the MFV 
hypothesis provides a solution 
to the flavour problem.

\subsubsection{General considerations}

The idea that the CKM matrix rules the strength of 
FCNC transitions also beyond the SM is a concept
that has been implemented and discussed in several works, 
especially after the first results of the $B$ factories
(see e.g.~Ref.~\cite{MFV2,Buras:2003jf}). 
However, it is worth stressing that the CKM matrix 
represents only one part of the problem: a key role in
determining the structure of FCNCs is also played  by quark masses, or 
better by the Yukawa eigenvalues. 
In this respect, the above MFV criterion provides the maximal protection 
of FCNCs (or the minimal violation of flavour symmetry), since the full 
structure of Yukawa matrices is preserved. Moreover, 
contrary to other approaches, the above MFV criterion 
is based on a renormalization-group-invariant symmetry argument,
which can easily be extended to TeV-scale 
effective theories where new degrees of freedoms, 
such as extra Higgs doublets or SUSY partners of the SM fields
are included. 

In particular, it is worth stressing that the MFV hypothesis 
can be implemented also in the so-called Higgsless models,
i.e.~assuming that the breaking of the electroweak symmetry is 
induced by some strong dynamics at the TeV scale (similarly 
tot he breaking of chiral symmetry in QCD). In this case 
Eq.~(\ref{eq:SMY}) is replaced by an effective interaction 
between fermion fields and the Goldstone bosons associated 
to the spontaneous breaking of the gauge symmetry.
From the point of view of the flavour symmetry, this interaction 
is identical to the one in (\ref{eq:SMY}), and this allows us 
to proceed as in  the case with the explicit Higgs field
 (see e.g.~Ref.~\cite{Barbieri:2008zt}).
The only difference between weakly- and strongly-interacting 
theories at the TeV scale is that in the latter case the 
expansion in powers of the Yukawa spurions is not 
necessarily a rapidly convergent series. If this is the case, 
then a resummation of the terms involving the top-quark Yukawa 
coupling needs to be performed~\cite{Feldmann:2008ja}

This model-independent structure does not hold in 
most of the alternative definitions of MFV models 
that can be found in the literature. For instance, 
the definition of Ref.~\cite{Buras:2003jf} 
(denoted constrained MFV, or CMFV)
contains the additional requirement that the 
effective FCNC operators playing a significant 
role within the SM are the only relevant ones 
also beyond the SM. 
This condition is realized only in weakly coupled 
theories at the TeV scale with only one light Higgs doublet, 
such as the MSSM with small $\tan\beta$.
It does not hold in several other frameworks, such as 
Higgsless models, or the MSSM with large $\tan\beta$. 

\medskip

Although the MFV seems to be a natural solution to the flavor problem, 
it should be stressed that we are still
very far from having proved the validity of this hypothesis from 
data. A proof of the MFV hypothesis 
can be achieved only with a positive evidence of physics beyond 
the SM exhibiting the flavor-universality pattern
(same relative correction in $s\to d$, $b\to d$, and  $b\to s$ transitions
of the same type) predicted by the MFV assumption. 
While this goal is quite difficult to be achieved, 
the MFV framework is quite predictive and thus could 
easily be falsified. Some of the most interesting predictions
which could be tested in the near future are the following:
\begin{itemize}
\item{} No new CPV phases in $B_s$ mixing, hence $|\phi_{B_s}| <0.05$ 
from~$\cA_{\rm CP}(B_s \to \psi \phi)$. 
\item{} Ratio of  $B_s$ and $B_d$ decays into $\ell^+\ell^-$ pairs
determined by the CKM matrix:  $\cB(B_d\to \ell^+\ell^-)/\cB(B_s\to \ell^+\ell^-) \approx |V_{td}/V_{ts}|^2$.
\item{} No new CPV phases in $b\to s \gamma$, hence vanishingly small CP asymmetries in 
$B\to K^* \gamma$ and $B\to K^* \ell^+\ell^-$.
\end{itemize}
Violations of these bounds would not only imply 
physics beyond the SM, but also a clear signal of 
new sources of flavour symmetry breaking beyond the
Yukawa couplings.

\subsubsection{MFV at large $\tan\beta$.}
\label{eq:largetanb}

If the Yukawa Lagrangian contains more than one Higgs field,
we can still assume that the Yukawa couplings are the only 
irreducible breaking sources of ${\mathcal G}_{q}$, but 
we can change their overall normalization.
A particularly interesting scenario
is the two-Higgs-doublet model where 
the two Higgses are coupled separately to up-
and down-type quarks:
\begin{equation}
\mathcal{L}^{2HDM}_{Y}  =   {\bar Q}_L \lambda_d D_R  \phi_D
+ {\bar Q}_L Y_u U_R  \phi_U
+ {\bar L}_L Y_e E_R  \phi_D {\rm ~+~h.c.}
\label{eq:LY2}
\end{equation}
This Lagrangian is invariant under an extra ${U}(1)$
symmetry with respect to the one-Higgs Lagrangian 
in Eq.~(\ref{eq:SMY}): a symmetry under which 
the only charged fields are $D_R$ and $E_R$ 
(charge $+1$) and  $\phi_D$ (charge $-1$).
This symmetry, denoted ${U}_{\rm PQ}$, 
prevents tree-level FCNCs and implies that $Y_{u,d}$ are the only 
sources of ${\mathcal G}_{q}$ breaking appearing in the Yukawa
interaction (similar to the one-Higgs-doublet
scenario). Coherently with the MFV hypothesis, we can then 
assume that $Y_{u,d}$ are the only relevant 
sources of ${\mathcal G}_{q}$ breaking appearing 
in all the low-energy effective operators. 
This is sufficient to ensure that flavour-mixing 
is still governed by the CKM matrix, and naturally guarantees
a good agreement with present data in the $\Delta F =2$
sector. However, the extra symmetry of the Yukawa interaction allows
us to change the overall normalization of 
$Y_{u,d}$ with interesting phenomenological consequences
in specific rare modes. 

The normalization of the  Yukawa couplings is controlled
by the ratio of the vacuum expectation values  of the two Higgs fields, 
or by the parameter $\tan\beta = \langle \phi_U\rangle/\langle \phi_D\rangle=v_u/v_d$.
Defining the eigenvalues $\lambda_{u,d}$ as in Eq.~(\ref{eq:Ydbasis}),
\bea
\lambda_u &=& \frac{1}{v_u} {\rm diag}(m_u,m_c,m_t)~, \no \\
\lambda_d &=& \frac{1}{v_d} {\rm diag}(m_d,m_s,m_b) = \frac{\tan\beta}{v_u}
{\rm diag}(m_d,m_s,m_b).
\eea
For $\tan\beta >\!\! > 1 $ the smallness of the $b$ quark can be attributed to the smallness 
of of $v_d$ with respect to $v_u \approx v$, rather than to the smallness of the 
corresponding Yukawa coupling.
As a result, for $\tan\beta >\!\! > 1$ we cannot anymore neglect 
down-type Yukawa couplings. Since the $b$-quark Yukawa coupling 
becomes $\mathcal{O}(1)$, the large-$\tan\beta$ regime is particularly 
interesting for all the helicity-suppressed observables in $B$ physics
(i.e.~the observables suppressed within the SM by the smallness of the 
 $b$-quark Yukawa coupling).

Another important aspect of this scenario is that the 
the ${U}(1)_{\rm PQ}$ symmetry cannot be exact:
it has to be broken at least in the scalar potential
in order to avoid the presence of a massless pseudoscalar Higgs boson.
Even if the breaking of  ${U}(1)_{\rm PQ}$  and 
 ${\mathcal G}_{q}$ are decoupled, the presence of 
${U}(1)_{\rm PQ}$ breaking sources can have important 
implications on the  structure of the Yukawa interaction,
especially if $\tan\beta$ is large~\cite{Hall:1993gn,Babu:1999hn}.
{}
We can indeed consider new dimension-four operators such as
\begin{equation}
 \epsilon~ {\bar Q}_L  \lambda_d D_R  \tilde \phi_U
\qquad {\rm or} \qquad 
 \epsilon~ {\bar Q}_L  \lambda_u\lambda_u^\dagger \lambda_d D_R  \tilde \phi_U~,
\label{eq:O_PCU}
\end{equation}
where $\epsilon$ denotes a generic MFV-invariant
${U}(1)_{\rm PQ}$-breaking source. Even if $\epsilon \ll 1 $, 
the product $\epsilon \times  \tan\beta$ can be $\mathcal{O}(1)$, 
inducing large corrections to the down-type Yukawa 
sector:
\begin{equation}
 \epsilon~ {\bar Q}_L  \lambda_d D_R  \tilde \phi_U
\ \stackrel{vev}{\longrightarrow}  \
 \epsilon~  {\bar Q}_L  \lambda_d D_R   \langle \phi_U \rangle = 
  (\epsilon\times\tan\beta)~  {\bar Q}_L  \lambda_d D_R   \langle \phi_D \rangle~.
\end{equation}
This is what happens in supersymmetry, where the operators in Eq.~(\ref{eq:O_PCU}) are
generated at the one-loop level [$\epsilon \sim 1/(16\pi^2)$], and the
large $\tan\beta$ solution is particularly welcome in the contest of 
Grand Unified models~\cite{Olechowski:1988gh}.

One of the clearest phenomenological 
consequences is a suppression (typically in the $10-50\%$ range)
of the $B \to \ell \nu$ decay
rate with respect to its SM expectation~\cite{Hou:1992sy}.
{} 
But the most striking signature could arise from the 
rare decays $B_{s,d}\to \ell^+\ell^-$
whose rates could be enhanced over the SM expectations 
by more than one order of magnitude~\cite{Hamzaoui:1998nu} 
{}
as already shown in Fig.~\ref{fig:tbbsmumu} in the context of supersymmetric models. 
An enhancement of both $B_{s}\to \ell^+\ell^-$ and 
$B_{d}\to \ell^+\ell^-$ respecting the MFV relation
$\Gamma(B_{s}\to \ell^+\ell^-)/\Gamma(B_{d}\to \ell^+\ell^-)
\approx  F^2_{B_s}/F^2_{B_d} |V_{ts}/V_{td}|^2$ would be an unambiguous signature 
of MFV at large $\tan\beta$~\cite{Blanke:2006ig,Hurth:2008jc}. 

\subsection{Flavour breaking in the MSSM}
\label{sect:SUSY}

The Minimal Supersymmetric extension of the SM (MSSM) 
is one of the most well-motivated and definitely the most 
studied extension of the SM at the TeV scale. For a detailed 
discussion of this model we refer to the review in Ref.~\cite{Martin:1997ns}
and to the lectures by J.~Ellis at this school.
Here we limit our self to analyse some properties 
of this model relevant to flavor physics. 

The particle content of the MSSM consists of the SM gauge and 
fermion fields plus a scalar partner for each quark and lepton 
(squarks and sleptons) and a spin-1/2 partner for each 
gauge field (gauginos). The Higgs sector has two Higgs
doublets with the corresponding spin-1/2 partners (higgsinos)
and a Yukawa coupling of the type in Eq.~(\ref{eq:LY2}).
While gauge and Yukawa interactions of the model are 
completely specified in terms of the corresponding SM
couplings, the so-called soft-breaking sector\footnote{~Supersymmetry 
must be broken in order to be consistent with observations
(we do not observe degenerate spin partners in nature). The 
soft breaking terms are the most general supersymmetry-breaking 
terms which preserve the nice ultraviolet properties of the 
model. They can be divided into two main classes: 
1) mass terms which break the mass degeneracy of the  
spin partners (e.g.~sfermion or gaugino mass terms); 
ii) trilinear couplings among the scalar fields 
of the theory (e.g.~sfermion-sfermion-Higgs couplings).}
of the theory contains several new free parameters, most of which are 
related to flavor-violating observables. For instance the 
$6\times6$ mass matrix of the up-type squarks, 
after the up-type Higgs field gets a vev ($\phi_U \to \langle \phi_U \rangle$), 
has the following structure
\begin{equation}
{\tilde M}_U^2 = 
\left(
\begin{array}{cc}
  {\tilde m}_{Q_L}^2   &  A_U \langle \phi_U \rangle \\
  A_U^\dagger \langle \phi_U \rangle & {\tilde m}_{U_R}^2 
\end{array}
\right)~  + ~\mathcal{O}\left( m_Z, m_{\rm top} \right)~,
\end{equation}
where ${\tilde m}_{Q_L}$, ${\tilde m}_{U_R}$, and $A_U$ are $3\times3$ 
unknown 
matrices.
Indeed the adjective {\em minimal} in the MSSM acronyms refers 
to the particle content of the model but does not specify 
its flavor structure. 

Because of this large number of free parameters, we cannot 
discuss the implications of the MSSM in flavor physics 
without specifying in more detail the flavor structure of the model. 
The versions of the MSSM analyzed in the literature range from 
the so-called Constrained MSSM (CMSSM), where the complete model 
is specified in terms of only four free parameters (in addition to the 
SM couplings), to the MSSM without $R$ parity and generic flavor 
structure, which contains a few hundreds of new free parameters.

Throughout the large amount of work in the past decades it has 
became clear that the MSSM with generic flavor structure and 
squarks in the TeV range is not compatible with precision tests 
in flavor physics. This is true even if we impose $R$ parity, 
the discrete symmetry which forbids single s-particle production, 
usually advocated to prevent a too fast proton decay. In this 
case we have no tree-level FCNC amplitudes, but the loop-induced
contributions are still too large compared to the SM ones 
unless the squarks are highly degenerate or have very small
intra-generation mixing angles. This is nothing but a 
manifestation in the MSSM context of the general flavor problem
illustrated in the first lecture.

The flavor problem of the MSSM is an important clue about the underling 
mechanism of supersymmetry breaking. On general grounds, mechanisms 
of SUSY breaking with flavor universality (such as gauge mediation) 
or with heavy squarks (especially in the case of the 
first two generations) tends to be favored. However, several options are 
still open.
These range from the very restrictive CMSSM case, which is a 
special case of MSSM with MFV, to more general scenarios with 
new small but non-negligible sources of flavor symmetry breaking.

\subsubsection{Flavor Universality, MFV, and RGE in the MSSM.}
\label{sect:MSSM_MFV}
Since the squark fields have well-defined transformation
properties under the SM quark-flavor group ${\mathcal G}_q$,
the MFV hypothesis can easily be implemented in the MSSM 
framework following the general rules outlined in 
Sect.~\ref{sect:MFV}. 

We need to consider all possible interactions compatible 
with i) softly-broken supersymmetry; ii) the breaking of 
${\mathcal G}_q$ via the spurion fields $Y_{U,D}$. 
This allows to express the squark mass terms and 
the trilinear quark-squark-Higgs couplings 
as follows~\cite{Hall:1990ac,D'Ambrosio:2002ex}:
\begin{eqnarray}
{\tilde m}_{Q_L}^2 &=& {\tilde m}^2 \left( a_1 {1 \hspace{-.085cm}{\rm l}} 
+b_1 Y_U Y_U^\dagger +b_2 Y_D Y_D^\dagger 
+b_3 Y_D Y_D^\dagger Y_U Y_U^\dagger +\ldots
 \right)~, \nonumber  \\
{\tilde m}_{U_R}^2 &=& {\tilde m}^2 \left( a_2 {1 \hspace{-.085cm}{\rm l}} 
+b_5 Y_U^\dagger Y_U +\ldots \right)~,  \qquad 
\nonumber\\
A_U &=&~ A\left( a_3 {1 \hspace{-.085cm}{\rm l}} 
+b_6 Y_D Y_D^\dagger +\ldots \right) Y_U~,\qquad 
\label{eq:MSSMMFV}
\end{eqnarray}
and similarly for the down-type terms. 
The dimensional parameters $\tilde m$ and $A$,
expected to be in the range few 100 GeV -- 1 TeV, 
set the overall scale of the soft-breaking terms.
In Eq.~(\ref{eq:MSSMMFV}) we have explicitly shown  
all independent flavor structures which cannot be absorbed into 
a redefinition of the leading terms (up to tiny contributions 
quadratic in the Yukawas of the first two families),  
when $\tan\beta$ is not too large and the bottom Yukawa coupling 
is small, the terms quadratic in $Y_D$ can be dropped.

In a bottom-up approach, the dimensionless coefficients 
$a_i$ and $b_i$ should be considered as free parameters 
of the model. Note that  this structure is 
renormalization-group invariant: the values of 
$a_i$ and $b_i$ change according to the 
Renormalization Group (RG) flow, but the general structure 
of Eq.~(\ref{eq:MSSMMFV})
is unchanged. This is not the case if the $b_i$ are set to zero,
corresponding to the so-called hypothesis of {\em flavor universality}.
In several explicit mechanism of supersymmetry breaking, 
the condition of flavor universality
holds at some high scale $M$, such as the scale of 
Grand Unification in the CMSSM (see below) or the mass-scale of the
messenger particles in gauge mediation (see Ref.~\cite{Giudice:1998bp}). 
In this case  non-vanishing 
$b_i \sim (1/4\pi)^2 \ln M^2/ {\tilde M}^2$ are 
generated by the RG evolution. 
As recently 
pointed out in Ref.~\cite{Paradisi:2008qh}
{} the RG flow in the MSSM-MFV 
framework exhibit quasi infra-red fixed points: even
if we start with all the $b_i =\mathcal{O}(1)$ at some high scale, 
the only non-negligible terms at the TeV scale are those 
associated to the $Y_U Y_U^\dagger$ structures.

If we are interested only in low-energy processes we can integrate 
out the supersymmetric particles at one loop and project this 
theory into the general MFV effective theory approach discussed before.
In this case the coefficients of the dimension-six effective operators 
written in terms of SM and Higgs fields
are computable in terms of the supersymmetric soft-breaking parameters.
The typical effective scale suppressing these operators 
(assuming an overall coefficient $1/\Lambda^2$) is
$\Lambda \sim  4 \pi\tilde m$. Since the bounds on $\Lambda$ 
within MFV are in the few TeV range, we then conclude that  
if MFV holds, the present bounds on FCNCs do not exclude squarks in 
the few hundred GeV mass range, i.e.~well within the LHC reach.

\subsubsection{The CMSSM framework.}
\label{sect:CMSSM}

The CMSSM, also known as mSUGRA, is the supersymmetric extension 
of the SM with the minimal particle content and the maximal 
number of universality conditions on the soft-breaking terms. 
At the scale of Grand Unification ($M_{\rm GUT} \sim 10^{16}$~GeV) 
it is assumed that there are only three independent 
soft-breaking terms: the universal gaugino mass (${\tilde m}_{1/2}$), 
the universal trilinear term ($A$), and the universal 
sfermion mass ($\tilde m_0$). The model has two additional 
free parameters in  the Higgs sector (the so-called $\mu$ 
and $B$ terms), which control the vacuum expectation 
values of the two Higgs fields (determined also by the 
RG running from the unification scale down to the electroweak scale). 
Imposing the correct $W$- and $Z$-boson masses allow us 
to eliminate one of these Higgs-sector parameters, the remaining 
one is usually chosen to be $\tan\beta$. As a result, the model is fully 
specified in terms of the three high-energy parameters
$\{ {\tilde m}_{1/2}, {\tilde m}_0, A \}$, and the 
low-energy parameter $\tan\beta$.\footnote{More precisely, 
for each choice of $\{ {\tilde m}_{1/2}, {\tilde m}_0, A, \tan\beta\}$
there is a discrete ambiguity related to the sign of the $\mu$ term.}
This constrained version of the MSSM is an example of a SUSY
model with MFV. Note, however, that the model is much more 
constrained than the general MSSM with MFV: in addition to be 
flavor universal, the soft-breaking terms at the unification 
scale obey various additional constraints 
(e.g.~in Eq.~(\ref{eq:MSSMMFV}) we have $a_1=a_2$ and $b_i=0$).

\begin{figure}[t]
\begin{center}
  \resizebox{7cm}{!}{\includegraphics{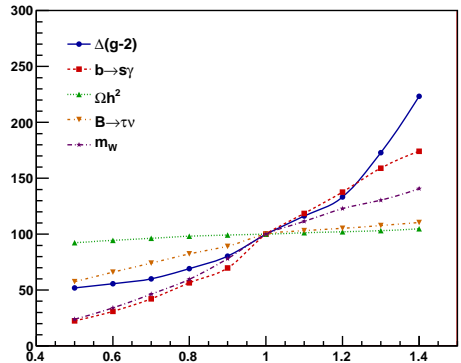}}
\end{center}
\caption {Relative sizes of the 95\%~C.L.\ preferred area in 
  the $(m_0, \tan\beta)$ plane of the CMSSM, as a function
  hypothetical variations of the present errors 
  on $(g-2)_\mu$, the cold dark-matter abundance ($\Omega h^2$),
  $\cB(B \to X_s \gamma)$, $\cB(B^+ \to \tau^+ \nu)$, and  $m_W$~\cite{Buchmueller:2007zk}.
  The error scaling is relative to the current combined theory and experimental
  error. }
\label{fig:drill}
\end{figure}

In the MSSM with $R$ parity we can distinguish 
five main classes of one-loop diagrams contributing 
to FCNC and CP violating processes with external down-type 
quarks. They are distinguished according to the virtual particles running 
inside the loops: $W$ and up-quarks (i.e.~the leading SM amplitudes),
charged-Higgs and up-quarks, charginos and up-squarks, 
neutralinos and down-squarks, gluinos and down-squarks. Within the CMSSM, 
the charged-Higgs and chargino exchanges yield the dominant 
non-standard contributions. 

Given the low number of free parameters, the CMSSM is very predictive 
and phenomenologically constrained by the precision measurements 
in flavor physics. The most powerful low-energy constraint comes 
from $B \to X_s \gamma$. For large values of $\tan \beta$, strong 
constraints are also obtained from $B_s \to \mu^+ \mu^-$, 
and $B^+ \to \tau^+ \nu$. If these observables 
are within the present experimental bounds, the constrained nature 
of the model implies essentially no observable deviations from the SM 
in other flavor-changing processes.  Interestingly enough, the CMSSM 
satisfy at the same time the flavor constraints and those from 
electroweak precision observables for squark masses below 1 TeV~\cite{Buchmueller:2007zk}.
An illustration of the constraining power of the flavour 
observables in determining the allowed parameter space of the 
model is shown in Fig.~\ref{fig:drill}. As can be noted, a reduction
or the error on $B \to X_s \gamma$ by 50\% would imply a substantial 
reduction of the allowed region in the CMSSM parameter space.

It is worth to stress that as long as we relax the strong  universality
assumptions of the CMSSM, the phenomenology of the model can
vary substantially. An illustration of this statement is provided by 
 Fig.~\ref{fig:tbbsmumu}, where we compare the predictions for 
 $B_s \to \mu^+ \mu^-$ in the CMSSM and in its  minimal variation,
the so-called Non-Universal Higgs Mass (NUHM) scenario. In the latter 
case only the condition of universality for the soft breaking terms in the 
Higgs sectors is relaxed, increasing by one unit the number of free parameters
of the model. As can be noted, the difference is substantial (in both cases
all existing constraints are satisfied). This also illustrate how 
precise data from flavour physics are essential to discriminate different 
versions of the MSSM.  
A recent detailed analysis of the discriminating power of 
flavour observables with respect to different versions of
the MSSM, even beyond MFV, can be found in Ref.~\cite{Altmannshofer:2009ne}.

\subsubsection{The Mass Insertion Approximation in the general MSSM.}
\label{sect:genFCNC}
Flavor universality at the GUT scale is not a 
general property of the MSSM, even if the model is embedded 
in a Grand Unified Theory. If this assumption is relaxed, 
new interesting phenomena can occur in flavor physics. 
The most general one is the appearance of gluino-mediated
one-loop contributions to FCNC 
amplitudes~\cite{Ellis:1981ts}

The main problem when going beyond simplifying assumptions, 
such as flavor universality or MFV, is the proliferation in 
the number of free parameters. A useful model-independent 
parametrization to describe the new phenomena occurring 
in the general MSSM with R parity conservation 
is the so-called mass insertion (MI) 
approximation~\cite{Hall:1985dx}. Selecting a flavor basis 
for fermion and sfermion states  where all the couplings 
of these particles to neutral gauginos are flavor diagonal,
the new flavor-violating effects are parametrized in terms 
of the non-diagonal entries of the sfermion mass matrices. 
More precisely, denoting by $\Delta$ the off-diagonal terms in the
sfermion mass matrices (i.e. the mass terms relating sfermions of the
same electric charge, but different flavor), the sfermion propagators
can be expanded in terms of $\delta = \Delta/ \tilde{m}^2$,
where $\tilde{m}$ is the average sfermion mass.  As long as $\Delta$
is significantly smaller than $\tilde{m}^2$ (as suggested by the absence 
of sizable deviations form the SM), one can truncate the series 
to the first term of this expansion and the experimental information
concerning FCNC and CP violating phenomena translates into upper
bounds on these $\delta$'s~\cite{Gabbiani:1996hi}.

The major advantage of the MI method is that it is not necessary to 
perform a full diagonalization of the sfermion mass matrices, 
obtaining a substantial simplification in the comparison of 
flavor-violating effects 
in different processes. There exist four type of mass insertions 
connecting
flavors $i$ and $j$ along a sfermion propagator:
$\left(\Delta_{ij}\right)_{LL}$, $\left(\Delta_{ij}\right)_{RR}$,
$\left(\Delta_{ij}\right)_{LR}$ and
$\left(\Delta_{ij}\right)_{RL}$. The indexes $L$ and $R$ refer to the
helicity of the fermion partners.

In most cases the leading non-standard amplitude 
is the gluino-exchange one, which is enhanced by one or two powers of 
the ratio $(\alpha_{\rm strong}/\alpha_{\rm weak})$ with respect 
to neutralino- or chargino-mediated amplitudes. When analyzing the 
bounds, it is customary to consider one non-vanishing MI at a time, 
barring accidental cancellations.  This procedure is
justified a posteriori by observing that the MI bounds have
typically a strong hierarchy, making the destructive interference
among different MIs rather unlikely. The bound thus obtained 
from recent measurements in $B$ and $K$ 
physics are reported in 
Tab.~\ref{tab:MIquarks}.
\begin{table}
  \centering
  \begin{tabular}{lll} 
    $\vert(\delta^d_{12})_{LL,RR}\vert < 1\cdot 10^{-2} $\
    &
    $\vert( \delta^d_{12} )_{LL=RR} \vert < 2 \cdot 10^{-4}$\
    &
    $\vert( \delta^d_{12})_{LR}\vert < 5\cdot 10^{-4} $\
\\ \hline
    $\vert(\delta^d_{13} )_{LL,RR} \vert < 7\cdot 10^{-2}$\
    &
    $\vert( \delta^d_{13})_{LL=RR}\vert < 5 \cdot 10^{-3}$\
    &
    $\vert( \delta^d_{13} )_{LR}\vert < 1\cdot 10^{-2}$  \
\\ \hline
    $\vert( \delta^d_{23} )_{LL}\vert < 2\cdot 10^{-1}$\
    &
    $\vert( \delta^d_{23} )_{LL=RR} \vert <  5\cdot 10^{-2}$\
    &
    $\vert(  \delta^d_{23} )_{LR,RL}\vert < 5 \cdot 10^{-3}$ \
\\ \\
\end{tabular}
\caption{Upper bounds at 95\% C.L.~on the dimensionless 
down-type mass-insertion parameters (see text) 
for squark and gluino masses of 350 GeV (from 
Ref.~\cite{Buchalla:2008jp}). } 
\label{tab:MIquarks}
\end{table}
The bounds mainly depend on the gluino and on the average 
squark mass, scaling as the inverse mass (the inverse mass 
square) for bounds derived from $\Delta F=2$ ($\Delta F=1$)
observables. 

The only clear pattern emerging from these bounds 
is that there is no room for sizable 
new sources of flavor-symmetry breaking. 
However, it is too early to draw
definite conclusions since some of the bounds,
especially those in the 2-3 sector,
are still rather weak. As suggested by various authors
(see eg. Ref.~\cite{Ciuchini:2007ha}), the possibility 
of sizable deviations from the SM in the 2-3 
sector could fit well with the large 2-3 mixing 
of light neutrinos, in the context of 
a unification of quark and lepton 
sectors. The could possibly explain a large CP violating
phase in $B_s$--$\bar B_s$ mixing.

\subsection{Flavour protection from hierarchical fermion profiles}
 
So far we have assumed that the suppression of flavour-changing
transitions beyond the SM can be attributed to a flavour symmetry, 
and a specific form of the symmetry-breaking terms. An interesting 
alternative is the possibility of a generic {\em  dynamical suppression}
of flavour-changing interactions, related to the weak mixing of 
the light SM fermions with the new dynamics at the TeV scale. 
A mechanism of this type is the so-called RS-GIM mechanism
occurring in models with a warped extra dimension.
In this framework the hierarchy of fermion masses, which 
is attributed  to the different localization of the fermions in 
the bulk~\cite{AS}, implies that the lightest fermions 
are those localized far from the infra-red (SM) brane. 
As a result, the suppression of FCNCs involving 
light quarks is a consequence of the small overlap of the light 
fermions with the lightest Kaluza-Klein excitations~\cite{aps}. 

As pointed out in~\cite{Davidson:2007si}, 
also the general features of this class 
of models can be described by means of an effective theory approach.
The two main assumptions of this approach are
the following: 
\begin{itemize}
\item{}
There exists a (non-canonical) basis 
for the SM fermions where their kinetic terms exhibit 
a rather hierarchical form: 
\bea
&& {\cal L}^{\rm quarks}_{\rm kin} = \sum_{\Psi=Q_L, U_R, D_R }
 \overline{\Psi} Z_\psi^{-2}   D\sla \Psi~, \no\\
&& Z_\psi =  {\rm diag}(z_\psi^{(1)}, z_\psi^{(2)}, z_\psi^{(3)}  )~, 
\qquad  z_\psi^{(1)}\ll z_\psi^{(2)}\ll z_\psi^{(3)} \lsim 1~.
\eea
\item{} In such basis there  is no flavour symmetry and all the 
flavour-violating interactions, including the Yukawa 
couplings, are $\cO(1)$. 
\end{itemize}
Once the fields are transformed 
into the canonical basis, the hierarchical kinetic terms 
act as a distorting lens, through which all interactions 
are seen as approximately aligned on the magnification axes 
of the lens. As anticipated, this construction provide
an effective four-dimensional description of a wide class of 
models with a with a warped extra dimension.
However, it should be stresses that this mechanism is not 
a general feature of models with extra dimensions: as discussed 
in~\cite{Cacciapaglia:2007fw}, 
also in extra-dimensional models is possible to postulate the 
existence of additional symmetries and, for instance, recover 
a MFV structure. 

The dynamical mechanism of hierarchical fermion profiles 
is quite effective in suppressing 
FCNCs beyond the SM. In particular, it can be shown 
that all the dimensions-six FCNC left-left operators
(such as the $\Delta F=2$ terms in Eq.~(\ref{eq:dfops})), 
have the same suppression as in MFV~\cite{Davidson:2007si}.
However, a residual problem is present in the 
left-right operators contributing to CP-violating 
observables in the kaon 
system: $\epsilon_K$~\cite{Bona:2007vi,Csaki1}
{}  and $\epsilon^\prime/\epsilon_K$~\cite{noi},
with potentially visible effects also in 
rare $K$ decays~\cite{Blanke:2008yr}.
As a result, contrary to most of the models discussed 
before, in this framework one expects no
significant NP effects in the $B$ system, 
while possible improved predictions in the $K$ system 
could reveal some deviation from the SM.

\section{Conclusions} 
      
The absence of significant deviations from the SM 
in quark flavour physics is a key information about any 
extension of the SM.  Only models with a highly non 
generic flavour structure can both stabilize the 
electroweak sector and, at the same time, 
be compatible with flavour observables. 
In such models we expect new particles within the LHC reach; however,
the structure of the new theory cannot be
determined using only the high-$p_T$ data from LHC.
As illustrated in these lectures, there are still various
open questions about the flavour structure of the 
model that can be addressed only at low energies, and
in particular with $B$ decays.

The set of $B$-physics observables to be measured 
with higher precision, and the rare transitions to be searched for 
is limited, if we are interested only on physics beyond the SM.
But is far from being a small set. As shown in these lectures, 
we know very little yet about CP violation the the $B_s$
system, and about FCNC transitions of the type $B \to K^* \ell^+\ell^-$ 
and $B_{s,d} \to \ell^+\ell^-$. And a systematic reduction in the 
determination of the SM Yukawa couplings, such as the 
determination of $\gamma$ from $B\to DK$ decays, could possibly reveal 
non-standard effects also in observables which we have already 
measured well, such as $\epsilon_K$ or the $B_d$ mixing phase.

\section*{Acknowledgments}
I wish to thank the organizers of {\em SUSSP65} for the invitation 
to this interesting and well-organized school, as well as 
all the participants who contributed to create a very
enjoyable and stimulating atmosphere.

\end{document}